\documentstyle[aps,epsfig]{revtex}
\begin{document}

\vskip 6.2cm

\begin{center}

{\Large  Fractality and geometry in ultra-relativistic nuclear
collisions}
\vskip 1.2cm

I.Zborovsk\'{y}
\vskip 0.5cm
Nuclear Physics Institute,
Academy of Sciences of the Czech Republic,
\newline
250 68 \v{R}e\v{z}, Czech Republic

\vskip 2.0cm

Abstract
\end{center}
\vskip 0.4cm

Assuming fractality of hadronic constituents,
we argue that asymmetry of space-time can be induced in the
ultra-relativistic interactions of hadrons and nuclei.
The asymmetry is expressed in terms of the
anomalous fractal dimensions of the colliding objects.
Besides state of motion, the relativistic principle is applied to the
state of asymmetry as well.
Such realization of relativity
concerns scale dependence of physical laws emerging at small distances.
We show that induced asymmetries of space-time
are a priori not excluded by the Michelson's experiment
even at large scales.

\vskip 0.5cm

\noindent
PACS number(s): 03.30.+p, 05.45.Df, 13.85.Ni

\newpage

{\section{Introduction}}
At sufficient high energies, the interactions of hadrons and nuclei
can be considered as an ensemble of individual interactions
of their constituents.
The constituents are partons in the parton model or
quarks and gluons which are the building blocks in the theory of QCD.
Production of particles with large transverse momenta
from such reactions has relevance to
fundamental principles of physics at small interaction distances.
One of the expressions aiming to reflect this relation is the $z$
scaling \cite{Z99}
observed in differential cross sections for the inclusive
reactions at high energies.
The hypothesis of the scaling states that
cross sections of particles with large transverse momentum $Q_{\bot}$
produced in relativistic collisions of hadrons and
nuclei depend on the single variable
\begin{equation}
z = z_0\Omega^{-1},
\label{eq:u1}
\end{equation}
where

\begin{equation}
\Omega(x_1,x_2) = (1-x_1)^{\delta_1}(1-x_2)^{\delta_2}.
\label{eq:u2}
\end{equation}
The scaling variable $z$ has character of a fractal
measure \cite{Mandelbrot}. For a given production process,
its finite part $z_0$ is proportional to the transverse
energy released in the underlying collision of constituents.
The divergent part $\Omega^{-1}$
describes the resolution at which the  collision of
the constituents can be singled out of this process.
The $\Omega(x_1,x_2)$
represents relative number of all initial
configurations containing the constituents which carry the
fractions $x_1$ and $x_2$ of the incoming momenta.
In such a picture,
we consider the ultra-relativistic collisions of hadrons and nuclei as
collisions of parton fractals with the anomalous fractal
dimensions $\delta_1$ and $\delta_2$.

The goal of the paper is to focus on the general premisses
of the $z$ scaling in view of fundamental principles of physics
at small interaction distances. It concerns scale dependence
of physical laws gradually emerging in various experimental and
theoretical investigations \cite{Wheeler,Hawking}.
Such extension of physics is intrinsically linked to the
evolution of the concept of space-time
\cite{Ord,Nottale,Pissondes}.
Its structure is characterized by explicitly
scale dependent metric potentials. Asking questions about
the metrics leads one to question the relativity. The
relativistic principle
besides motion, applies also to the laws of scale.
The basic assumption tackled in the paper is breaking of the
reflection invariance which is the characteristic feature of
fractality at small distances. One of the consequences is
change of the standard dispersion relation between the energy and
momentum having implications to non-standard expressions of
these quantities in terms of the velocity.
New dispersion relation is Lorenz transformation of the
free energy with respect to change of the state of scale. The scale
changes are expressed in terms of the "scale velocity" which
is property of space-time induced by the interaction. We express
the scale velocity connected with the space-time asymmetry in
terms of the anomalous fractal dimensions of the interacting
fractal objects. The geometrical objects model the internal
parton structure of hadrons and nuclei revealed in their
interactions at high energies. Dealing with asymmetry of
space-time, we show that anisotropic propagation of light
signals need not to be necessary in contradiction with
Michelson's experiment concerning the light interference.

\vskip 0.5cm
{\section{Constituent interactions}}

Ultra-relativistic interactions of hadronic constituents
are local relative to the resolution which depends on the
kinematical characteristics of particles produced in the
collisions.
In accordance with the property of locality it has been suggested
\cite{Stavinsky} that gross features of the
single-inclusive particle distributions for the reaction

\begin{equation}
M_1+M_2 \rightarrow m_1 + X
\label{eq:u3}
\end{equation}
can be described in terms of the corresponding kinematical
characteristics of the interaction

\begin{equation}
(x_1M_1) + (x_2M_2) \rightarrow m_1 +
(x_1M_1+x_2M_2 + m_2).
\label{eq:u4}
\end{equation}
The $M_1$ and $M_2$ are masses of the colliding
hadrons or nuclei and $m_1$ is the mass of the inclusive
particle.
The parameter $m_2$ is used in connection with internal
conservation laws (for isospin, baryon number, and strangeness).
The $x_{1}$ and $x_{2}$ are the scale-invariant fractions of the
incoming four-momenta $P_{1}$ and $P_{2}$ of the colliding objects.
We have determined the momentum fractions $x_1$ and $x_2$ in the way to
minimize the fractal resolution $\Omega^{-1}(x_1,x_2)$
accounting simultaneously for the recoil mass condition

\begin{equation}
(x_1P_1 + x_2P_2 - Q)^{2} = (x_1M_1 + x_2M_2 + m_2)^{2} .
\label{eq:u5}
\end{equation}
The $Q$  is the four-momentum of the inclusive particle
with the mass $m_1$. The momentum fractions resulting
from these requirements represent the sum

\begin{equation}
x_1 = \lambda_1 + \chi_1 , \ \ \ \ \ \ \
x_2 = \lambda_2 + \chi_2 .
\label{eq:u6}
\end{equation}
The parts $\lambda_i$ are connected with the inclusive particle
and the parts $\chi_i$ with the creation of its recoil.
According to the decomposition, the binary subprocess
(\ref{eq:u4}) can be rewritten to the symbolic form

\begin{equation}
(\lambda_1+\chi_1) + (\lambda_2+\chi_2) \rightarrow
(\lambda_1+\lambda_2)+(\chi_1+\chi_2).
\label{eq:u7}
\end{equation}
The explicit formulae for the momentum fractions $\lambda_i$ read

\begin{equation}
\lambda_1 = \frac{(P_{2}Q)+M_2m_2}{(P_{1}P_{2})-M_1M_2} ,\ \ \ \ \
\lambda_2 = \frac{(P_{1}Q)+M_1m_2}{(P_{1}P_{2})-M_1M_2}.
\label{eq:u8}
\end{equation}
Their combinations

\begin{equation}
\sqrt{\lambda_1\lambda_2} \sim \frac{E_{\bot}}{\sqrt{s}}, \ \ \ \ \ \ \
\sqrt{\frac{\lambda_2}{\lambda_1}} \sim \tan(\theta/2)
\label{eq:u9}
\end{equation}
are related to the transverse energy $E_{\bot}$
and the centre-of-mass angle $\theta$
of the inclusive particle $m_1$.
Now let us turn to the recoil part in the constituent subprocess
and examine it in more detail.
Suppose both colliding objects posses similar
fractal structures in the sense that their anomalous fractal dimensions
are equal $\delta_1=\delta_2$. In that case, the momentum fractions
$\chi_1$ and $\chi_2$ take the form

\begin{equation}
\chi_1=\overline{\mu}_1 \equiv
\overline{\mu} \sqrt{\frac{1-\lambda_1}{1-\lambda_2}}, \ \ \ \ \ \ \
\chi_2=\overline{\mu}_2 \equiv
\overline{\mu} \sqrt{\frac{1-\lambda_2}{1-\lambda_1}},
\label{eq:u10}
\end{equation}
where
\begin{equation}
\overline{\mu} \equiv \sqrt{\lambda_1\lambda_2+\lambda_0}, \ \ \ \ \ \ \
\lambda_0 = \frac{0.5(m_2^{2}-m_1^{2})}{(P_{1}P_{2})-M_1M_2} .
\label{eq:u11}
\end{equation}
Similar as for the inclusive particle, the combinations

\begin{equation}
\sqrt{\overline{\mu}_1\overline{\mu}_2}
\sim \frac{E_{\bot}'}{\sqrt{s}}, \ \ \ \ \ \ \
\sqrt{\frac{\overline{\mu}_2}{\overline{\mu}_1}}
\sim \tan(\overline{\theta}/2)
\label{eq:u12}
\end{equation}
are related to the transverse energy $E_{\bot}'$ and the centre-of-mass
angle $\overline{\theta}$
of the recoil in the binary subprocess (\ref{eq:u4}).
The center-of-mass angels do not
comply the back-to-back correlation $\theta+\overline{\theta}=\pi$
between the inclusive particle and its recoil for $x_1\ne x_2$.
This is because the center-of-mass system of the interacting
constituents is generally not equal to the center-of-mass system
of the reaction. The transverse energy balance is expressed by
the relation $\overline{\mu}^{2}=\overline{\mu}_1\overline{\mu}_2$.

Now suppose the colliding objects possess mutually different
fractal structures in the sense that their anomalous fractal dimensions
are not equal, $\delta_1\ne\delta_2$.
In that case, the momentum fractions $\chi_1$ and
$\chi_2$ have more complicated form

\begin{equation}
\chi_1 = \sqrt{\mu_1^{2}+\omega_1^{2}}-\omega_1 , \ \ \ \ \ \
\chi_2 = \sqrt{\mu_2^{2}+\omega_2^{2}}+\omega_2 ,
\label{eq:u13}
\end{equation}
where

\begin{equation}
\mu_1 = \overline{\mu}_1\sqrt{\alpha} ,
\ \ \ \ \ \ \ \ \
\mu_2 = \overline{\mu}_2\frac{1}{\sqrt{\alpha}} ,
\label{eq:u14}
\end{equation}
\begin{equation}
\omega_1 = \mu_1\overline{a} , \ \ \ \ \ \ \ \ \ \
\omega_2 = \mu_2\overline{a} .
\label{eq:u15}
\end{equation}
The parameter $\alpha = \delta_2/\delta_1$ is ratio of the
anomalous (fractal) dimensions of the fractal objects (hadrons
and nuclei) colliding at high energy. The symbol $\bar{a}$ is given by
the formula

\begin{equation}
\overline{a} = \frac{\alpha-1}{2\sqrt{\alpha}}\xi
\label{eq:u16}
\end{equation}
where

\begin{equation}
\xi = \frac{\overline{\mu}}{ \sqrt{ (1-\lambda_1)(1-\lambda_2)}}
\label{eq:u17}
\end{equation}
is a scale factor from the interval $0\le\xi\le 1$.
When approaching the phase-space
limit, the scale factor $\xi$ tends to unity. Along the phase-space
limit $\xi=1$ and $x_1=x_2=1$.
The phase-space boundary thus  corresponds to the fractal limit
characterized by infinite value of the fractal measure
(\ref{eq:u1}), i.e. by infinite value of the scaling variable $z$.
For collisions of the asymmetric objects with  $\delta_1\ne \delta_2$,
we denote the  center-of-mass angle of the recoil
in the subprocess (\ref{eq:u4})  by $\theta'$ .
The transverse energy $E_{\bot}'$ and the angle $\theta'$
correspond  to the following combinations of the
momentum fractions

\begin{equation}
\sqrt{\chi_1\chi_2} \sim \frac{E_{\bot}'}{\sqrt{s}}, \ \ \ \ \ \ \
\sqrt{\frac{\chi_2}{\chi_1}} \sim \tan(\theta'/2).
\label{eq:u18}
\end{equation}
The conservation of transverse degrees of freedom is given by
$\overline{\mu}^{2}=\chi_1\chi_2$.
It can be shown from the above relations that

\begin{equation}
\theta' \le \overline{\theta}   \ \ \ for \ \ \
\delta_1\le\delta_2
\label{eq:u19}
\end{equation}
and vice versa. We thus make the following conclusion.
In the collision of fractal objects with
mutually different anomalous dimensions, the momentum of the recoil
produced in a constituent collision is shifted towards the
fractal object with richer fractal structure expressed by larger anomalous
dimension. In this sense the interactions of constituents
are influenced by the asymmetric  fractal background created in
collisions of parton fractals (hadrons or nuclei) with
$\delta_1\ne \delta_2$.
The asymmetry of such background represents a sort
of "medium" with scale properties
characterized by a scale velocity $\nu$.

There exists analogy of such  situation which is the
propagation of light in moving refracting media. If a medium
with the refractive index $n$ moves with the velocity $v$,
the elementary waves are dragged along the medium with the
velocity \cite{Moller}

\begin{equation}
a= \frac{v}{1-v^2/n^2}\left(1-\frac{1}{n^{2}}\right).
\label{eq:u20}
\end{equation}
There is, however, significant difference between this analogy
and the situation we consider in ultra-relativistic nuclear
collisions. The difference is because of the basic property of
fractals - never ending structure at any resolution.
Basic assumption here
is that fractality of the interaction distorts
the very structure of space-time in the interaction region.
As a result the space-time becomes polarized with the metric
undergoing change.
Background metric changes have been considered as "recoil"
effects modifying the relativistic momentum-energy dispersion
relation \cite{Ellis}. The particle feels an 'unusual' metric
which is a constant of motion. Fractalization of space-time was
considered in Ref. \cite{Ord,Nottale} and its properties have been
studied by others \cite{Pissondes}. One of its basic properties
is breakdown of the reflection invariance which depends on scale.

Generally, fractal approach to the ultra-relativistic
interactions of hadrons and nuclei needs profound understanding.
It concerns the deformation of
space-time at small scales in the interaction region
and attributes additional meaning
to the physical quantities such as the momentum, mass, energy or
velocity. They may be defined from parameters of the fractal
objects in terms of the fractal geometry \cite {Nottale}.
This includes extension of the relativity principles to the
relativity of scales as well to more comprehensive
scale-motion relativistic concepts.

\vskip 0.5cm
{\section {Break down of the reflection invariance, the way towards
scale-motion relativity}}

General solution to the theory of the special relativity is the
Lorenz transformation. As demonstrated by Nottale, it can be
obtained under minimal number of three successive constraints.
They are (i) homogeneity of space-time translated as the linearity
of the transformation, (ii) the group structure defined
by the internal composition law and (iii) isotropy of
space-time expressed as the reflection invariance.
Let us consider the relativistic boost along the x-axis.
Without any loss of generality, the linearity of the
transformation can be expressed in the form

\begin{equation}
x'  = \gamma (u)[x-ut] ,
\label{eq:u21}
\end{equation}

\begin{equation}
t'  = \gamma (u)[A(u)t-B(u)x] ,
\label{eq:u22}
\end{equation}
where $\gamma$, $A$, and $B$ are functions of a parameter $u$.
The parameter represents usual velocity in the motion relativity
or the "scale velocity" used, e.g., in the concept of the scale
relativity concerning fractal dimensions and fractal
measures \cite{Nottale}.
The principle of relativity tells us that these equations keep
the same form whatever the state of motion. The third constraint,
the isotropy of space-time, results in the requirement that
change of orientation of the variable axis does not change the
form of the transformations, provided $u'=-u$.
As considered in the previous section, the fractal approach
to the ultra-relativistic
interaction of hadrons and nuclei leads to the space-time
isotropy breakdown in the interaction region.
This is translated as breaking of the reflection
invariance at the infinitesimal level \cite{Pissondes}.

\vskip 0.5cm
\subsection{Space-time asymmetry in 3+1 dimensions}

We now turn to the question how to express breaking of the
reflection invariance in the framework of special relativity.
Let us describe a point $P$ in two Cartesian reference
systems $S$ and $S'$.
We assume that the systems are oriented parallel to each other and that
$S'$ is moving relative to $S$ with the velocity $u$ in the
direction of the positive $x$-axis. We suppose that the
asymmetry expressed by a parameter $a$ is parallel to the
velocity $u$.
As demonstrated in Appendix,
the relativistic transformations of the coordinates
and time are given by

\begin{equation}
x_1' = \gamma(u) \left[x_1-ut\right] , \ \ \ \ \
t' = \gamma(u)
\left[\left(1-2au\right)t-ux_1\right] ,
\label{eq:u23}
\end{equation}
where

\begin{equation}
\gamma(u) = \frac{1}{\sqrt{1-2au-u^{2}}}.
\label{eq:u24}
\end{equation}
The violation of the space-time reflection invariance is
expressed by a non-zero value of $a$.
For the vanishing value of $a$, the transformations turn into the usual
relativistic transformations of the coordinates and time.
The inverse relations

\begin{equation}
x_1 = \gamma(u) \left[(1-2au)x_1'+ut'\right] ,
\ \ \ \ \ \ \
t = \gamma(u) \left[t'+ux_1'\right]
\label{eq:u25}
\end{equation}
are obtained as the solution of Eqs. (\ref{eq:u23}) and (\ref{eq:u24})
with respect to the unprimed variables.
They can be also derived by
the interchange $x_1\leftrightarrow x_1'$,
$t\leftrightarrow t'$, $u\leftrightarrow u'$, and by the relation

\begin{equation}
u' = -\frac{u}{1-2au} .
\label{eq:u26}
\end{equation}
This formula connects the velocity $u'$ of the system $S$ in the
$S'$ frame with the velocity $u$ of the system $S'$ in the $S$
reference system.
Because of the asymmetry parameter $a$, the magnitudes of the two
velocities are not equal.
The invariant of the transformations (\ref{eq:u23}) is

\begin{equation}
t^{2}-x^{2}_1-2tax_1 .
\label{eq:u27}
\end{equation}
In more general case, when the space-time anisotropy
$\bbox{a}$ acquires an arbitrary direction, we write
the invariant in the form

\begin{equation}
\eta_{\mu\nu}(\bbox{a})x^{\mu}x^{\nu} =
t^{2}-\bbox{x}^{2}-
2t\bbox{a}\!\cdot\!\bbox{x} - (\bbox{a}\!\times\!\bbox{x})^2
\equiv \tau^2 .
\label{eq:u28}
\end{equation}
Besides the diagonal part, it has extra
terms given by a non-zero values of the vector $\bbox{a}$.
Similar extra terms of the relativistic invariant were considered in
Ref. \cite{Pissondes} and associated with breaking of the reflection
invariance assumed at the infinitesimal level.
In the four dimensional notation,
the invariant (\ref{eq:u28}) corresponds to the metrics

\begin{equation}
\eta_{\mu\nu}(\bbox{a}) =
\left(
\begin{array}{cc}
\eta_{ij}  & -a_i \\
-a_j & 1 \\
\end{array}
\right), \ \ \ \ \ \ \ \
\eta_{ij}=-(1+a^2)\delta_{ij}+a_ia_j.
\label{eq:u29}
\end{equation}
Here the indices $i$ and $j$ numerate the first three rows and
columns of the matrix $\eta$, respectively.
The $\delta_{ij}$ is the Kronecker's symbol.
Next we present the explicit form for the relativistic
transformations of the coordinates and time in the considered case.
They must be linear and homogeneous, preserving
the invariant (\ref{eq:u28}).
The transformations have to possess an internal group structure
required by the principle of relativity.
We denote the parameter of the group by the symbol $\bbox{u}$.
The parameter is the velocity
of the system $S'$ with respect to the $S$ reference frame.
In connection with the transformation formulae, it is convenient
to introduce the notations
\begin{equation}
\gamma = \frac{1}
{\sqrt{1-u^2-
2\bbox{a}\!\cdot\!\bbox{u} - (\bbox{a}\!\times\!\bbox{u})^2}}
\label{eq:u30}
\end{equation}
and
\begin{equation}
g = (1-\bbox{a}\!\cdot\!\bbox{u})\gamma-1 .
\label{eq:u31}
\end{equation}
Here $a^{2}=\bbox{a}\!\cdot\!\bbox{a}$ and
$u^{2}=\bbox{u}\!\cdot\!\bbox{u}$.
We define the following combinations of $g$ and $\gamma$,
\begin{equation}
\gamma_{\pm} = g \pm \gamma\bbox{a}\!\cdot\!\bbox{u} ,
\ \ \ \ \ \ \ \
g_{\pm} =
\gamma (1+a^{2}) \pm
g \bbox{a}\!\cdot\!\bbox{u}/u^{2} .
\label{eq:u32}
\end{equation}
Let us consider the relativistic transformations
\begin{equation}
\bbox{x}' = \bbox{x} -
\bbox{u}\left[\gamma(t-\bbox{a}\!\cdot\!\bbox{x}) -
g\bbox{u}\!\cdot\!\bbox{x}/u^{2}\right] ,
\label{eq:u33}
\end{equation}
\begin{equation}
t' = t +
\left[\gamma_{-}(t-\bbox{a}\!\cdot\!\bbox{x}) -
g_{-}\bbox{u}\!\cdot\!\bbox{x}\right] .
\label{eq:u34}
\end{equation}
They generalize the special transformations (\ref{eq:u23})
which are recovered by
$\bbox{u} = (u,0,0)$ and $\bbox{a} = (a,0,0)$.
The inverse relations are obtained
by the interchange $\bbox{x} \leftrightarrow \bbox{x}'$,
$t \leftrightarrow t'$, $\bbox{u} \leftrightarrow \bbox{u}'$,
where
\begin{equation}
\bbox{u}' = -\frac{\bbox{u}}{1-2\bbox{a}\!\cdot\!\bbox{u}} .
\label{eq:u35}
\end{equation}
According to the substitution, there exist the symmetry properties
\begin{equation}
\gamma(\bbox{u}') =
(1-2\bbox{a}\!\cdot\!\bbox{u})\gamma(\bbox{u}) , \ \ \ \ \
\gamma_{\pm}(\bbox{u}') = \gamma_{\mp}(\bbox{u}) ,
\label{eq:u36}
\end{equation}
\begin{equation}
g(\bbox{u}') = g(\bbox{u}) , \ \ \ \ \ \
g_{\pm}(\bbox{u}') =
(1-2\bbox{a}\!\cdot\!\bbox{u})g_{\mp}(\bbox{u}) .
\label{eq:u37}
\end{equation}
Exploiting the properties, the inverse transformations
\begin{equation}
\bbox{x} = \bbox{x}' +
\bbox{u}\left[\gamma(t'-\bbox{a}\!\cdot\!\bbox{x}') +
g\bbox{u}\!\cdot\!\bbox{x}'/u^{2}\right] ,
\label{eq:u38}
\end{equation}
\begin{equation}
t = t' +
\left[\gamma_{+}(t'-\bbox{a}\!\cdot\!\bbox{x}') +
g_{+}\bbox{u}\!\cdot\!\bbox{x}'\right] .
\label{eq:u39}
\end{equation}
follow immediately.
The relativistic transformations can be expressed
in a more compact form

\begin{equation}
x' = D(\bbox{u},\bbox{a})x ,
\label{eq:u40}
\end{equation}
where

\begin{equation}
D(\bbox{u},\bbox{a}) =
\left(
\begin{array}{cc}
\delta_{ij}\!+\!g u_iu_j/u^{2}\!+\!\gamma u_ia_j &
-\gamma u_i \\
-g_{-}u_j\!-\!\gamma_{-}a_j & 1\!+\!\gamma_{-} \\
\end{array}
\right) .
\label{eq:u41}
\end{equation}
The inverse matrix reads

\begin{equation}
D^{-1}(\bbox{u},\bbox{a}) =
\left(
\begin{array}{cc}
\delta_{ij}\!+\!g u_iu_j/u^{2}\!-\!\gamma u_ia_j &
+\gamma u_i \\
+g_{+}u_j\!-\!\gamma_{+}a_j & 1\!+\!\gamma_{+} \\
\end{array}
\right) .
\label{eq:u42}
\end{equation}
The transformation matrices can be decomposed into the product

\begin{equation}
D(\bbox{u},\bbox{a}) =
A^{-1}_x(\bbox{a}) \Lambda(\bbox{\beta}) A_x(\bbox{a}) .
\label{eq:u43}
\end{equation}
Here

\begin{equation}
A_x (\bbox{a}) =
\left(
\begin{array}{cc}
\sqrt{1+a^2}\delta_{ij} &  0 \\
-a_j &  1 \\
\end{array}
\right)
\label{eq:u44}
\end{equation}
and

\begin{equation}
\Lambda (\bbox{\beta}) =
\left(
\begin{array}{cc}
\delta_{ij}\!+\!g_{0} \beta_i\beta_j/\beta^{2} &
-\gamma_{0}\beta_i  \\
-\gamma_{0}\beta_j  &  \gamma_{0}\\
\end{array}
\right)  ,
\label{eq:u45}
\end{equation}
with

\begin{equation}
\gamma_{0} = \frac{1}{\sqrt{1\!-\!\beta^{2}}} , \ \ \ \ \ \ \ \ \
g_{0} = \gamma_{0}-1.
\label{eq:u46}
\end{equation}
The matrix $\Lambda$ depends on the vector

\begin{equation}
\bbox{\beta} =
\sqrt{1\!+\!a^2}\frac{\bbox{u}}{1-\bbox{a}\!\cdot\!\bbox{u}} .
\label{eq:u47}
\end{equation}
Let us notice that the interchange
$\bbox{u}\leftrightarrow\bbox{u}'$ is equivalent to the symmetry
$\bbox{\beta}\leftrightarrow -\bbox{\beta}$.
The relativistic transformations (\ref{eq:u40}) preserve the
invariant (\ref{eq:u28}). This follows from the relation

\begin{equation}
D^{\dag}(\bbox{u},\bbox{a}) \eta(\bbox{a}) D(\bbox{u},\bbox{a}) =
\eta(\bbox{a}) = A^{\dag}_x(\bbox{a})\eta_0 A_x(\bbox{a}) ,
\label{eq:u48}
\end{equation}
where $\eta_0$ stands for the diagonal matrix
$\eta_0$=diag(-1,-1,-1,+1).

The transformations comply the principle of relativity.
Mathematically it is expressed by their group properties.
Let $D(\bbox{u},\bbox{a})$ and $D(\bbox{v}',\bbox{a})$ be two
successive relativistic transformations represented by the
matrices (\ref{eq:u41}).
The composition of the  transformations has the form

\begin{equation}
\Omega_x(\bbox{\phi},\bbox{a}) D(\bbox{v},\bbox{a})  =
D(\bbox{v}',\bbox{a}) D(\bbox{u},\bbox{a}) ,
\label{eq:u49}
\end{equation}
provided

\begin{equation}
\bbox{v} = \frac{\bbox{v}' +
\bbox{u}\left[\gamma(1-\bbox{a}\!\cdot\!\bbox{v}')
+g\bbox{u}\!\cdot\!\bbox{v}'/u^{2}\right]  }
{1+\gamma_{+}(1-\bbox{a}\!\cdot\!\bbox{v}')
+g_{+}\bbox{u}\!\cdot\!\bbox{v}'} .
\label{eq:u50}
\end{equation}
One can obtain the above relations by exploiting the
decomposition (\ref{eq:u43}) and
using the structure of the Lorenz group expressed by the
formula

\begin{equation}
R(\bbox{\phi}) \Lambda(\bbox{\beta}_v) =
\Lambda(\bbox{\beta}_{v'}) \Lambda(\bbox{\beta}_{u}) .
\label{eq:u51}
\end{equation}
The matrix

\begin{equation}
R(\bbox{\phi}) =
\left(
\begin{array}{cc}
r_{ij} & 0 \\
0 & 1  \\
\end{array}
\right) , \ \ \ \ \
\bbox{\phi} = \bbox{v}'\times\bbox{u}
\label{eq:u52}
\end{equation}
describes the Thomas precession \cite{Thomas} around the vector
$\bbox{\phi}$ known in the theory of relativity.
The angle of the precession $\varphi$ depends on the vectors
$\bbox{\beta}_{v'}$ and $\bbox{\beta}_u$.
It remains to identify

\begin{equation}
\Omega_x(\bbox{\phi},\bbox{a}) =
A^{-1}_x(\bbox{a}) R(\bbox{\phi}) A_x(\bbox{a})
\label{eq:u53}
\end{equation}
and we get Eq. (\ref{eq:u49}).
The relativistic transformation of the coordinates and
time with rotation of the coordinate axes has the structure

\begin{equation}
D(\bbox{u},\bbox{a}) \Omega_x(\bbox{\phi},\bbox{a}) =
\left(
\begin{array}{cc}
r_{ij}\!+\!g u_iu_kr_{kj}/u^{2}
\!+\!\gamma u_ia_j &
-\gamma u_j  \\
-g_{-}u_kr_{kj}\!+\!a_kr_{kj}\!-\!
(1\!+\!\gamma_{-}) a_j
 & 1\!+\!\gamma_{-} \\
\end{array}
\right) ,
\label{eq:u54}
\end{equation}
provided the asymmetry of space-time is expressed by the
vector $\bbox{a}$.
As concerns Eq. (\ref{eq:u50}), it can be obtained from the usual
relativistic composition of the factors $\bbox{\beta}$ given by
Eq. (\ref{eq:u47}).
The inverse relation

\begin{equation}
\bbox{v}' = \frac{\bbox{v} -
\bbox{u}\left[
\gamma(1-\bbox{a}\!\cdot\!\bbox{v})
-g\bbox{u}\!\cdot\!\bbox{v}/u^{2}\right]  }
{1+\gamma_{-}(1-\bbox{a}\!\cdot\!\bbox{v})-
g_{-}\bbox{u}\!\cdot\!\bbox{v}}
\label{eq:u55}
\end{equation}
corresponds to the composition of the transformations in the
following form

\begin{equation}
\Omega_x(-\bbox{\phi},\bbox{a}) D(\bbox{v}',\bbox{a}) =
D(\bbox{v},\bbox{a}) D^{-1}(\bbox{u},\bbox{a}) .
\label{eq:u56}
\end{equation}
When using  Eqs. (\ref{eq:u50}) and (\ref{eq:u55}),
we get

\begin{equation}
1-\bbox{a}\!\cdot\!\bbox{v} = \gamma
\frac{(1-\bbox{a}\!\cdot\!\bbox{u})
(1-\bbox{a}\!\cdot\!\bbox{v}')+
(1+a^{2})\bbox{u}\!\cdot\!\bbox{v}'}
{1+\gamma_{+}(1-\bbox{a}\!\cdot\!\bbox{v}')+
g_{+}\bbox{u}\!\cdot\!\bbox{v}'} ,
\label{eq:u57}
\end{equation}

\begin{equation}
1-\bbox{a}\!\cdot\!\bbox{v}'
 = \gamma
\frac{(1-\bbox{a}\!\cdot\!\bbox{u})(1-\bbox{a}\!\cdot\!\bbox{v})-
(1+a^{2})\bbox{u}\!\cdot\!\bbox{v} }
{1+\gamma_{-}(1-\bbox{a}\!\cdot\!\bbox{v})-
g_{-}\bbox{u}\!\cdot\!\bbox{v}}.
\label{eq:u58}
\end{equation}
It follows from the relations that

\begin{equation}
\gamma(\bbox{v}) =  \gamma(\bbox{v}')
\left[1+\gamma_{+}(1-\bbox{a}\!\cdot\!\bbox{v}')+
g_{+}\bbox{u}\!\cdot\!\bbox{v}'\right] ,
\label{eq:u59}
\end{equation}

\begin{equation}
\gamma(\bbox{v}') =  \gamma(\bbox{v})
\left[1+\gamma_{-}(1-\bbox{a}\!\cdot\!\bbox{v})-
g_{-}\bbox{u}\!\cdot\!\bbox{v}\right] .
\label{eq:u60}
\end{equation}
Region of the accessible values of the velocities is given by
the factor $\gamma$.
The boundary of the region is fixed by the condition
$\gamma(\bbox{v}) = \infty$.
For a given value of $\bbox{a}$, it is an ellipsoid

\begin{equation}
(v_{\parallel}+a)^2 + (1\!+\!a^2)v^2_{\bot} = 1+a^2
\label{eq:u61}
\end{equation}
in the velocity space.
The focus of the ellipsoid is situated into the point $\bbox{v}=0$.
The $v_{\parallel}$ and $v_{\bot}$ denote the velocity components
which are parallel and perpendicular to the vector
$\bbox{a}$, respectively.
The ellipsoid is invariant under the relativistic transformations
(\ref{eq:u50}) and (\ref{eq:u55}).
In the case of $\bbox{u}=(u,0,0)$ and $\bbox{a}=(a,0,0)$, the
composition of the velocities has the simple form

\begin{equation}
v_1' = \frac{v_1-u}{1-2au-uv_1}, \ \ \
v_i' = v_i\frac{\sqrt{1-2au-u^{2}}}{1-2au-uv_1}, \
\ \ \ \ \ i=2,3.
\label{eq:u62}
\end{equation}
The inverse relations can be obtained by
the interchange $\bbox{v}\leftrightarrow \bbox{v}'$ and
$u\leftrightarrow u'$. Using Eq. (\ref{eq:u26}),
they can be written as follows

\begin{equation}
v_1 = \frac{v_1'+u-2auv_1'}{1+uv_1'}, \ \ \
v_i = v_i'\frac{\sqrt{1-2au-u^{2}}}{1+uv_1'}, \
\ \ \ \ i=2,3.
\label{eq:u63}
\end{equation}

{\subsection {Energy and momentum}}

Consider a material particle in space-time.
In relativistic mechanics, the position and momentum of the
particle are given by the four-vectors
$x^{\mu}=\{\bbox{x},t\}$ and $p^{\mu}=\{\bbox{p},E\}$,
respectively.
Let us define an "elementary" particle as an object which
reveals no internal structure at any resolution considered.
We comprehend the notion of elementarity as a relative
concept which relies on the scales we are dealing with.
For the infinite resolution it should be a perfect
point whose trajectory is a fractal curve.
For an arbitrary small but still finite resolution,
the perfect point is approximated by a particle which we
call "elementary" with respect to this resolution.
It is therefore natural to assume that
the concepts of the momentum, energy,  mass and the velocity of the
"elementary" particle have good physical meaning also
at the scales where space-time is expected to break down its
isotropy.
We impose the following requirements on the energy and momentum.

1. The energy of a free particle cannot be pumped from the
structure of space-time. This condition reads

\begin{equation}
E = E_{min} \ \ \ \ \ \ for \ \ \ \ \ \bbox{v}=0
\label{eq:u64}
\end{equation}
where $\bbox{v}$ is velocity of the particle.

2. Rate of clocks is slowest in the centre of gravity system.
The only source of gravity is the free particle itself and
not the structure of space-time. This condition reads

\begin{equation}
dt = dt_{min} \ \ \ \ \ \ for \ \ \ \ \ \bbox{p}=0
\label{eq:u65}
\end{equation}
where $\bbox{p}$ is the momentum of a free particle.

3. The transverse mass expressed relative to the space-time
asymmetry $\bbox{a}$ is conserved. The requirement states that
energy of a particle moving perpendicular to the asymmetry
$\bbox{a}$ is equal to the transverse mass of the particle
given by the transverse component of the momentum $\bbox{p}_{\bot}$
with respect to $\bbox{a}$. This is

\begin{equation}
E^2 = p_{\bot}^2+m_0^2 \ \ \ \ \ \ for \ \ \ \ \
\bbox{v}=\bbox{v}_{\bot} .
\label{eq:u65a}
\end{equation}
The physical requirements are obvious in the Minkovski space-time. In the
case the asymmetry occurs, they lead to specific constraints on
the energy and momentum.
We denote the values of the momentum and the energy of a
free particle by ($\bbox{p}$ and  $E$) or
($\bbox{p}'$ and $E'$) in the reference systems $S$ or $S'$,
respectively. In consistence with the principle of relativity and
the ideas presented above, we search for relations connecting
these quantities.
In order to do that, let us first determine the 4-momentum
$\pi^{\mu}=\{\bbox{\pi},\pi_0\}$
with space component in the direction of the asymmetry
$\bbox{a}$.
The components of the variables determined relative to the systems
$S$ and $S'$ transform in the way

\begin{equation}
\pi' = \Pi(\bbox{u},\bbox{a})\pi ,
\label{eq:u66}
\end{equation}
where
\begin{equation}
\Pi (\bbox{u},\bbox{a}) = D^{\dag} (\bbox{u},-\bbox{a}) .
\label{eq:u67}
\end{equation}
The explicit form of the transformation matrix is

\begin{equation}
\Pi(\bbox{u},\bbox{a}) =
\left(
\begin{array}{cc}
\delta_{ij}\!+\!g u_iu_j/u^{2}\!-\!\gamma a_iu_j
& -g_{+}u_i\!+\!\gamma_{+}a_i \\
 -\gamma u_j  & 1\!+\!\gamma_{+} \\
\end{array}
\right) .
\label{eq:u68}
\end{equation}
The inverse transformation
is given by

\begin{equation}
\Pi^{-1} (\bbox{u},\bbox{a}) =
\left(
\begin{array}{cc}
\delta_{ij}\!+g u_iu_j/u^{2}\!+\!\gamma a_iu_j
& +g_{-}u_i\!+\!\gamma_{-}a_i \\
+\gamma u_j & 1\!+\!\gamma_{-} \\
\end{array}
\right)
\label{eq:u69} .
\end{equation}
According to the relation (\ref{eq:u67}),
the matrix $\Pi$ can be expressed in the form

\begin{equation}
\Pi(\bbox{u},\bbox{a}) = A^{-1}_{\pi}(\bbox{a})
\Lambda(\bbox{\beta}) A_{\pi}(\bbox{a})
\label{eq:u70}
\end{equation}
where

\begin{equation}
A^{-1}_{\pi}(\bbox{a}) = A^{\dag}_x(-\bbox{a}) .
\label{eq:u71}
\end{equation}
The group properties of the transformations (\ref{eq:u66}) are
given by the composition

\begin{equation}
\Omega_{\pi}(\bbox{\phi},\bbox{a}) \Pi(\bbox{v},\bbox{a}) =
\Pi(\bbox{v}',\bbox{a}) \Pi(\bbox{u},\bbox{a}) ,
\label{eq:u72}
\end{equation}
provided the velocities $\bbox{u}$, $\bbox{v}'$, and
$\bbox{v}$ satisfy the relation (\ref{eq:u50}).
Here

\begin{equation}
\Omega_{\pi} (\bbox{\phi},\bbox{a}) =
A^{-1}_{\pi}(\bbox{a}) R(\bbox{\phi}) A_{\pi}(\bbox{a}) .
\label{eq:u73}
\end{equation}
We show that Eqs. (\ref{eq:u49}) and (\ref{eq:u72})
are consistent with relation (\ref{eq:u67}).
Let us transpose Eq. (\ref{eq:u49}).
Exploiting the correspondence (\ref{eq:u67}) and using
Eqs. (\ref{eq:u53}), (\ref{eq:u71}), and (\ref{eq:u73}), we can write

\begin{equation}
\Pi(\bbox{v},\bbox{a}) \Omega_{\pi}(-\bbox{\phi},\bbox{a}) =
\Pi(\bbox{v},\bbox{a}) \Omega^{\dag}_x(\bbox{\phi},-\bbox{a}) =
\Pi(\bbox{u},\bbox{a}) \Pi(\bbox{v}',\bbox{a}) .
\label{eq:u74}
\end{equation}
We apply the transposition operation on Eq. (\ref{eq:u51})
too.
As the matrices $\Lambda$ are invariant under the operation, we
obtain the composition of the parameters $\bbox{\beta}_u$ and
$\bbox{\beta}_{v'}$ in the mutual reverse order. From the
symmetry reasons, the composition must be of the same
form as Eq. (\ref{eq:u51}). We have therefore

\begin{equation}
R(-\bbox{\phi}) \Lambda (\bbox{\beta}_w) =
\Lambda (\bbox{\beta}_v) R(-\bbox{\phi}) =
\Lambda (\bbox{\beta}_u) \Lambda (\bbox{\beta}_{v'}) .
\label{eq:u75}
\end{equation}
The vector $\bbox{\beta}_w$ corresponds to the velocity
$\bbox{w}$ according to Eq. (\ref{eq:u47}).
The velocity is given by the formula (\ref{eq:u50}) in which the
velocities $\bbox{u}$ and $\bbox{v}'$ are mutually interchanged.
Multiplying Eq. (\ref{eq:u75}) by the $A^{-1}_{\pi}$ from the
left and by the $A_{\pi}$ from the right, we get

\begin{equation}
\Omega_{\pi}(-\bbox{\phi},\bbox{a}) \Pi(\bbox{w},\bbox{a}) =
\Pi(\bbox{v},\bbox{a}) \Omega_{\pi}(-\bbox{\phi},\bbox{a}) .
\label{eq:u76}
\end{equation}
Together with Eq. (\ref{eq:u74}) one has

\begin{equation}
\Omega_{\pi}(-\bbox{\phi},\bbox{a}) \Pi(\bbox{w},\bbox{a}) =
\Pi(\bbox{u},\bbox{a}) \Pi(\bbox{v}',\bbox{a}) .
\label{eq:u77}
\end{equation}
After performing the interchange
$\bbox{u}\leftrightarrow\bbox{v}'$,
we obtain Eq. (\ref{eq:u72}). It was thus shown that the
composition of two successive transformations
of the variables $\pi$ follows from the composition of
the corresponding transformations of the coordinates and time,
provided their transformation matrices are
connected by the relation (\ref{eq:u67}).

Now we determine the momentum $\bbox{p}$ of a free particle which in
general is not parallel to the asymmetry $\bbox{a}$.
Definition of the momentum has to be in consistence with the
condition 3 that requires preservation of the transverse mass
defined by the momentum components perpendicular to the
asymmetry $\bbox{a}$.
There exists two sets of the variables
$p^{\mu}_s = \{\bbox{p}_s,E\}$, $s=L,R$
which comply the requirement.
They are determined by the relation

\begin{equation}
\pi = A_s(\bbox{a})p_s ,
\label{eq:u78}
\end{equation}
where

\begin{equation}
A_s(\bbox{a}) =
\left(
\begin{array}{cc}
\delta_{ij}\!\pm\!\varepsilon_{ijk} a_k & 0 \\
0 & 1 \\
\end{array}
\right) .
\label{eq:u79}
\end{equation}
Here $\varepsilon_{ijs}$ is
the Levi-Civita symbol.
The plus (in the next every upper) sign and the minus (in the
next every lower) sign corresponds to $s=L$ and $s=R$, respectively.
We attribute the first set of the variables ($s=L$) to the
particle which we call left-handed.
The second set ($s=R$) corresponds to the particle
with right-handed type of motion.
The explicit relations between the momenta $\bbox{p}_s$ and
$\bbox{\pi}$ read

\begin{equation}
\bbox{\pi} = \bbox{p}_s\pm\bbox{p}_s\!\times\!\bbox{a} , \ \ \ \ \ \ \
\bbox{p}_s =
\frac{\bbox{\pi}\mp\bbox{\pi}\!\times\!\bbox{a}+
(\bbox{a}\!\cdot\!\bbox{\pi})\bbox{a}}
{1+a^{2}} .
\label{eq:u80}
\end{equation}
In the context of these definitions, it is suitable to
determine the associative variables $x^{\mu}_s=\{\bbox{x}_s,x_0\}$,
$s=L,R$, to the coordinates and time by the formula

\begin{equation}
x_s = A_s(-\bbox{a})x .
\label{eq:u81}
\end{equation}
The quantity $\bbox{u}_s=d\bbox{u}_s/dx_0$ is the velocity of
a particle for the special case when the particle moves in the
direction of $\bbox{a}$.
It is related to the velocity $\bbox{u}$ as follows

\begin{equation}
\bbox{u}_s = \bbox{u}\pm\bbox{a}\!\times\!\bbox{u} , \ \ \ \ \ \ \
\bbox{u} =
\frac{\bbox{u}_s\mp\bbox{a}\!\times\!\bbox{u}_s+
(\bbox{a}\!\cdot\!\bbox{u}_s)\bbox{a}}
{1+a^{2}} .
\label{eq:u82}
\end{equation}
Exploiting the additional notations

\begin{equation}
h \equiv \frac{g}{1+(\bbox{a}\cdot\bbox{u}_s)^{2}/u^2_s},
\ \ \ \ \ \ \
h_{\pm} \equiv \frac{g_{\pm}}{1+a^{2}} ,
\label{eq:u83}
\end{equation}
the relativistic transformations
of the energy and momentum take the form

\begin{equation}
p'_s = \Delta(\bbox{u}_s,\bbox{a})p_s ,
\label{eq:u84}
\end{equation}
where

\begin{equation}
\Delta(\bbox{u},\bbox{a}) =
\left(
\begin{array}{cc}
\delta_{ij}\!+\!h u_iu_j/u^{2}\!-\!h_{-} a_iu_j
& -h_{+}u_i\!+\!ha_i \\
-\gamma u_j & 1\!+\!\gamma_{+} \\
\end{array}
\right) .
\label{eq:u85}
\end{equation}
The inverse matrix reads

\begin{equation}
\Delta^{-1}(\bbox{u},\bbox{a}) =
\left(
\begin{array}{cc}
\delta_{ij}\!+\!hu_iu_j/u^{2}\!+\!h_{+} a_iu_j
& h_{-}u_i\!+\!ha_i \\
+\gamma u_j & 1\!+\!\gamma_{-} \\
\end{array}
\right) .
\label{eq:u86}
\end{equation}
The transformation matrixes  can be written in the way

\begin{equation}
\Delta(\bbox{u}_s,\bbox{a}) =
A_{ps}^{-1}(\bbox{a}) \Lambda(\bbox{\beta}) A_{ps}(\bbox{a}) ,
\label{eq:u87}
\end{equation}
where

\begin{equation}
A_{ps}(\bbox{a})=A_{\pi}(\bbox{a})A_s(\bbox{a}) =
\frac{1}{\sqrt{1+a^2}}
\left(
\begin{array}{cc}
\delta_{ij}\!\pm\!\varepsilon_{ijk} a_k & -a_i \\
0 & \sqrt{1+a^2} \\
\end{array}
\right)
\label{eq:u88}
\end{equation}
and $\Lambda$ is given by Eq. (\ref{eq:u45}).

The transformations (\ref{eq:u84}) possess group properties.
Let us consider two successive transformations expressed by
the matrices
$\Delta(\bbox{u}_s,\bbox{a})$ and $\Delta(\bbox{u}_s',\bbox{a})$.
The resultant transformation is given by

\begin{equation}
\Omega_{ps} (\bbox{\phi},\bbox{a})\Delta(\bbox{v}_s,\bbox{a}) =
\Delta (\bbox{v}_s',\bbox{a})\Delta (\bbox{u}_s,\bbox{a}) ,
\label{eq:u89}
\end{equation}
provided

\begin{equation}
\bbox{v}_s = \frac{\bbox{v}_s' +
\bbox{u}_s\left[\gamma - h_{-}\bbox{a}\!\cdot\!\bbox{v}_s'
+ h\bbox{u}_s\!\cdot\!\bbox{v}_s'/u_s^{2}\right]  }
{1+\gamma_{+}-h\bbox{a}\!\cdot\!\bbox{v}_s'
+h_{+}\bbox{u}_s\!\cdot\!\bbox{v}_s' } .
\label{eq:u90}
\end{equation}
Formula (\ref{eq:u89}) is consequence of Eqs.
(\ref{eq:u51}) and (\ref{eq:u87}).
The matrix $\Omega_{ps}$ has the structure

\begin{equation}
\Omega_{ps}(\bbox{\phi},\bbox{a}) =
A^{-1}_{ps}(\bbox{a}) R(\bbox{\phi}) A_{ps}(\bbox{a}) =
A_s^{-1}(\bbox{a})\Omega^{\dag}_x (-\bbox{\phi},-\bbox{a}) A_s(\bbox{a}) .
\label{eq:u91}
\end{equation}
The inverse relation to Eq. (\ref{eq:u90}) reads

\begin{equation}
\bbox{v}_s' = \frac{\bbox{v}_s -
\bbox{u}_s\left[\gamma-h_{+}\bbox{a}\!\cdot\!\bbox{v}_s-
h\bbox{u}_s\!\cdot\!\bbox{v}_s/u_s^{2} \right]  }
{1+\gamma_{-}-h\bbox{a}\!\cdot\!\bbox{v}_s-
h_{-}\bbox{u}_s\!\cdot\!\bbox{v}_s} .
\label{eq:u92}
\end{equation}
The composition rules (\ref{eq:u90}) and (\ref{eq:u92})
are obtained by substituting Eq. (\ref{eq:u82}) into the formulae
(\ref{eq:u50}) and (\ref{eq:u55}), respectively.

The four-momenta and position four-vectors shall not be
treated on the same footing at small scales in the region where
the asymmetry of space-time is induced by the interaction.
For a non-zero value of the asymmetry $\bbox{a}$ there exist
two sets of the mechanical variables $p^{\mu}_s$, $s=L,R$
attributed to the kinematical variables $x^{\mu}$.
Single sets of the mechanical variables correspond to the
right-handed and left-handed types of motion, respectively.
Both of them have either positive or negative energy.
The sign of the energy is conserved in whatever reference frame.
The variables $x^{\mu}$ and $p^{\mu}_s$ posses different
transformation properties.
While the first obey the transformation formula (\ref{eq:u40}),
the later are transformed according to Eq. (\ref{eq:u84}).
In the special case, when the velocity $\bbox{u}$ is parallel to
the vector $\bbox{a}$, $\bbox{u}=\bbox{u}_s$ and the transformation
matrices take the simple form

\begin{equation}
D(\bbox{u},\bbox{a}) =
\left(
\begin{array}{cc}
\delta_{ij}\!+\!\gamma_{+} u_iu_j/u^{2} & -\gamma u_i \\
-\gamma u_j & 1\!+\!\gamma_{-} \\
\end{array}
\right) , \ \ \ \ \ \
\Delta(\bbox{u},\bbox{a}) =
\left(
\begin{array}{cc}
\delta_{ij}\!+\!\gamma_{-} u_iu_j/u^{2} & -\gamma u_i \\
-\gamma u_j & 1\!+\!\gamma_{+} \\
\end{array}
\right) .
\label{eq:u93}
\end{equation}
The inverse matrices read

\begin{equation}
D^{-1}(\bbox{u},\bbox{a}) =
\left(
\begin{array}{cc}
\delta_{ij}\!+\!\gamma_{-} u_iu_j/u^{2} & +\gamma u_i \\
+\gamma u_j & 1\!+\!\gamma_{+} \\
\end{array}
\right) , \ \ \ \ \ \
\Delta^{-1}(\bbox{u},\bbox{a}) =
\left(
\begin{array}{cc}
\delta_{ij}\!+\!\gamma_{+} u_iu_j/u^{2} & +\gamma u_i \\
+\gamma u_j & 1\!+\!\gamma_{-} \\
\end{array}
\right) .
\label{eq:u94}
\end{equation}
As follows from the relation

\begin{equation}
\Delta^{\dag}(\bbox{u}_s) \eta(-\bbox{a}) \Delta(\bbox{u}_s) =
\eta(-\bbox{a}) = (1\!+\!a^2)
A^{\dag}_{ps}(\bbox{a})\eta_0 A_{ps}(\bbox{a}),
\label{eq:u95}
\end{equation}
the relativistic transformations (\ref{eq:u84}) preserve
the expression

\begin{equation}
p^2 = \frac{1}{1\!+\!a^2}
\eta_{\mu\nu}(-\bbox{a})p^{\mu}p^{\nu} =
\frac{1}{1\!+\!a^2}\left[E^2-\bbox{p}^{2}+
2E\bbox{a}\!\cdot\!\bbox{p}-
(\bbox{a}\!\times\bbox{p})^2\right]
\equiv m_0^2 .
\label{eq:u96}
\end{equation}
The invariant  is equal to  $m_0^2$, which
is square of the rest mass of a particle in the non-fractal
and non-relativistic mechanics.
Equation (\ref{eq:u96}) implies the dependence of the total energy of a
particle on its momentum in the following way

\begin{equation}
E = \sqrt{1\!+\!a^2}{\cal E}-\bbox{a}\!\cdot\!\bbox{p} ,
\label{eq:u97}
\end{equation}
where the symbol

\begin{equation}
{\cal E}
= \sqrt{p^2+m^2_0}
\label{eq:u98}
\end{equation}
stands for  the particle's free energy. Here we do not consider
the minus sign before the square root relevant for anti-particles.
The energy  (\ref{eq:u97}) is positive for arbitrary
values of the asymmetry $\bbox{a}$ and the momentum $\bbox{p}$.
It has a single minimum
corresponding to the momentum and energy

\begin{equation}
\bbox{p}_0 = m_0\bbox{a} , \ \ \ \ \ \ \
E(\bbox{p}_0) = m_0 ,
\label{eq:u99}
\end{equation}
respectively.
Beyond the minimum, as the momentum increases, the energy tends to
infinity.
The energy $E$ consists of two terms. The first term
is the free energy multiplied by the factor
$(1+a^2)^{1/2}$. The second term,
$V=-\bbox{a}\!\cdot\!\bbox{p}$, plays the role of a potential
induced by the asymmetry of space-time.
The asymmetry does not violate the energy momentum conservation.
We demonstrate it on a closed system with the mass $m_0$
which splits into two parts.
Denoting the four-momenta of the decay products
by $p_1$ and $p_2$, one can write

\begin{eqnarray}
m^2_0 &=& (p_1+p_2)^2 =
m^2_1+m^2_2+2p_1p_2 \nonumber \\
&=&
\frac{1}{1\!+\!a^2}
\left[E^2_1-\bbox{p}^2_1+2E_1\bbox{a}\!\cdot\!\bbox{p}_1
     -(\bbox{a}\!\times\!\bbox{p}_1)^2\right] +
\frac{1}{1\!+\!a^2}
\left[E^2_2-\bbox{p}^2_2+2E_2\bbox{a}\!\cdot\!\bbox{p}_2
     -(\bbox{a}\!\times\!\bbox{p}_2)^2\right]
\nonumber \\
& &+
\frac{2}{1\!+\!a^2}
\left[E_1E_2-\bbox{p}_1\!\cdot\!\bbox{p}_2+
E_1\bbox{a}\!\cdot\!\bbox{p}_2+E_2\bbox{a}\!\cdot\!\bbox{p}_1
-(\bbox{a}\!\times\!\bbox{p}_1)(\bbox{a}\!\times\!\bbox{p}_2)
\right]
\nonumber \\
&=&
\frac{1}{1\!+\!a^2}\left[
(E_1+E_2)^2-(\bbox{p}_1+\bbox{p}_2)^2+
2(E_1+E_2)\bbox{a}\!\cdot\!(\bbox{p}_1+\bbox{p}_2)
-\left(\bbox{a}\!\times\!(\bbox{p}_1\!+\!\bbox{p}_2)\right)^2
\right].
\label{eq:u100}
\end{eqnarray}
We see that if the four-momenta $p_1$ and $p_2$ are characterized
by the invariant (\ref{eq:u96}), their sum $p_1+p_2$ possesses
this property too.
This implies the conservation of the total energy and momentum,
$E=E_1+E_2$ and $\bbox{p}=\bbox{p}_1\!+\!\bbox{p}_2$,
which results in the conservation of the free energy

\begin{equation}
\sqrt{p^2+m^2_0} =
\sqrt{p^2_1+m^2_1} + \sqrt{p^2_2+m^2_2}
\label{eq:u101}
\end{equation}
as well.

\vskip 0.5cm
{\subsection {Relations of the kinematical and mechanical
variables}}

Fundamental concepts of the special theory of relativity lead us
to the relation between the energy/momentum of a
particle and its velocity. The velocity is limited within the sphere
of the radius $c=1$ in every system of reference and is oriented in
the direction of the particle momentum. This concerns the
homogeneous and isotropic space-time.
We show how the relations change when we abandon the space-time isotropy
which we expect to break down at small scales.

Let us first determine how depends the four-momentum
$\pi^{\mu}$ on the velocity $\bbox{v}$. According to the
definition (\ref{eq:u66}), the space component of $\pi^{\mu}$ is
parallel to the asymmetry $\bbox{a}$.
We search for the functions $\bbox{f}_1$ and $f_2$,

\begin{equation}
\bbox{\pi} = \bbox{f}_1(\bbox{v},\bbox{a}), \ \ \ \ \
\pi_0  = f_2(\bbox{v},\bbox{a}),
\label{eq:u102}
\end{equation}
which are form invariant with respect to the relativistic
transformations of the four-momentum $\pi^{\mu}$ and the velocity $\bbox{v}$.
One can convince itself that the expressions

\begin{equation}
\bbox{\pi} =
\left[(1+a^{2})\bbox{v} +
(1\!-\!\bbox{a}\!\cdot\!\bbox{v})\bbox{a}\right]
\gamma(\bbox{v})m_0 ,
\label{eq:u103}
\end{equation}

\begin{equation}
\pi_0  = (1\!-\!\bbox{a}\!\cdot\!\bbox{v})
\gamma(\bbox{v})m_0
\label{eq:u104}
\end{equation}
fulfill the requirements. Really, substituting the expressions
into the transformation formula (\ref{eq:u66}), one arrives at
the system

\begin{eqnarray}
\lefteqn{
\left(
\begin{array}{c}
(1\!+\!a^2)v_i' + (1\!-\!\bbox{a}\!\cdot\!\bbox{v}')a_i   \\
1-\bbox{a}\!\cdot\!\bbox{v}'  \\
\end{array}
\right)\gamma(\bbox{v}') =
}
\\ & &
\hspace*{-5mm} =\left(
\begin{array}{cc}
\delta_{ij}\!+\!g u_iu_j/u^2\!-\!\gamma a_iu_j
& -g_{+}u_i\!+\!\gamma_{+}a_i \\
-\gamma u_j & 1\!+\!\gamma_{+} \\
\end{array}
\right)
\left(
\begin{array}{c}
(1\!+\!a^2)v_i + (1\!-\!\bbox{a}\!\cdot\!\bbox{v})a_i  \\
1-\bbox{a}\!\cdot\!\bbox{v}  \\
\end{array}
\right)\gamma(\bbox{v})
\nonumber
\end{eqnarray}
consisting of four equations. The last one is identity.
This can be shown by using Eqs. (\ref{eq:u58}) and
(\ref{eq:u60}). In the same manner one can convince itself, that
the first three equations of the system are consistent with Eq.
(\ref{eq:u55}).
Now we substitute Eq. (\ref{eq:u78}) into the formulae
(\ref{eq:u103}) and (\ref{eq:u104}) and get

\begin{equation}
\bbox{p}_s = m\left(\bbox{v}\!+\!\bbox{a}\right)
\pm m(\bbox{a}\!\times\!\bbox{v}) ,
\label{eq:u106}
\end{equation}

\begin{equation}
E = m(1\!-\!\bbox{a}\!\cdot\!\bbox{v}) .
\label{eq:u107}
\end{equation}
These are the expressions for the momentum and the energy
of a left handed ($s\!=\!L$ with upper sign) and right handed
($s\!=\!R$ with lower sign) "elementary" particle moving
with the velocity $\bbox{v}$ in space-time characterized
by the vector anisotropy $\bbox{a}$.
As can be seen by direct calculation, the formulae
are consistent with the invariant (\ref{eq:u28}) and
(\ref{eq:u96}).
The coefficient of the proportionality
between the momentum $\bbox{p}_s$ and the velocity $\bbox{v}$
is denoted by the symbol $m$ and
represents the inertial mass of the particle.
The inertial mass depends on the velocity in the way

\begin{equation}
m = m_0\gamma(\bbox{v}) ,
\label{eq:u108}
\end{equation}
where $m_0$ is the rest mass of the particle
corresponding to the minimal energy (\ref{eq:u99}).
Another relation which is invariant under the
transformations (\ref{eq:u40}) and (\ref{eq:u84}) is action of
the free particle

\begin{eqnarray}
S_{as}  = -\tau m_0 =
-tE+\bbox{x}\!\cdot\!\bbox{p}_s
\pm\bbox{a}\!\cdot\!(\bbox{x}\!\times\!\bbox{p}_s) .
\label{eq:u109}
\end{eqnarray}
In classical mechanics, it determines the particle trajectory
expressed in terms of its momentum, energy, and the mass $m_0$.
The trajectory is given by $\bbox{x} = \bbox{v}t$,
where

\begin{equation}
\bbox{v} = \frac{\bbox{p}_s\pm\bbox{p}_s\!\times\!\bbox{a}-\bbox{a}E}
{E+\bbox{a}\!\cdot\!\bbox{p}_s}.
\label{eq:u110}
\end{equation}
If we substitute this expression into Eq. (\ref{eq:u109})
and exploit the formulae (\ref{eq:u97}), (\ref{eq:u106}), and
(\ref{eq:u107}), we get

\begin{equation}
t=\tau \gamma .
\label{eq:u111}
\end{equation}
The proportionality relates  the time $t$ recorded in the
observer's system $S$ to the particle's proper time $\tau$.
We see from Eq. (\ref{eq:u110})
that for the zero value of the momentum there exists the
non-zero value of the velocity

\begin{equation}
\bbox{v}_0 = -\bbox{a}
\label{eq:u112}
\end{equation}
which minimizes the relation (\ref{eq:u111}). This gives
the slowest rate of clocks in the center of gravity system
$\bbox{p}=0$ in consistence with the requirement 2.
As concerns the requirement 1, the minimal energy
(\ref{eq:u99}) corresponds to $\bbox{v}=0$.

Having determined the action $S_{as}$, we re-examine the Klein-Gordon
equation for free particle in space-time with the
asymmetry $\bbox{a}$. It follows from Eq. (\ref{eq:u40})
that the derivation operator
$\partial = (\bbox{\partial},\partial_0) $ transforms in the way

\begin{equation}
\partial' = (D^{\dag})^{-1}\partial .
\label{eq:u113}
\end{equation}
The Dalambertian operator is modified as  follows

\begin{equation}
\Box_{a} = \partial^{\dag}\eta^{-1}\partial .
\label{eq:u114}
\end{equation}
Its invariance  with respect
to the relativistic transformations is seen from the relation

\begin{equation}
\Box_{a} = \partial^{\dag}\eta^{-1}\partial
= \partial^{'\dag}\eta^{-1}\partial' = \Box'_{a} .
\label{eq:u115}
\end{equation}
Here we have exploited the decomposition

\begin{equation}
\eta^{-1}= D \eta^{-1}D^{\dag} .
\label{eq:u116}
\end{equation}
The explicit form of the operator (\ref{eq:u114}) reads

\begin{equation}
\Box_{a} = \partial_0^{2}-\frac{1}{1\!+\!a^2}
\left[\bbox{\partial}+\bbox{a}\partial_0\right]^{2} .
\label{eq:u117}
\end{equation}
The corresponding modified Klein-Gordon equation

\begin{equation}
-\Box_{a} \psi_{as} = m_0^2\psi_{as}
\label{eq:u118}
\end{equation}
has the solution

\begin{equation}
\psi_{as} = \exp(iS_{as}),
\label{eq:u119}
\end{equation}
where $S_{as}$ is the action given by Eq. (\ref{eq:u109}).
After inserting this solution into Eq. (\ref{eq:u118}), one arrives at the
equation (\ref{eq:u96}) relating the energy and momentum of a free
particle in space-time with the asymmetry $\bbox{a}$.

According to our opinion, the parameter could have relevance to
more deeper context of the metric potentials which have relation
to the intimate structure of space-time. It may be connected with
a "field of the space-time asymmetries" reflecting its structure
at small scales. Existence of such a "field" would result into a
disparity between the energy-momentum and the coordinates and
time. Here the disparity is demonstrated by the commutation
relation

\begin{equation}
A^{\dag}_{ps}\eta_0A_x - A^{\dag}_x\eta_0A_{ps} =
\left(
\begin{array}{cc}
\pm2\varepsilon_{ijk}a_k  & 0 \\
0 & 0 \\
\end{array}
\right)
\label{eq:u119a}
\end{equation}
between the matrices representing fluctuations in momenta and
coordinates, respectively.
The commutator is non-zero provided the non-zero value of the
"field". As a first step, one can approximate the "field
of space-time asymmetries" in terms of the asymmetry vector
$\bbox{a}$ and consider it as a random and chaotic quantity.
The investigations in this direction require, however, more
detailed and fundamental study.

\vskip 0.5cm
{\subsection {Relativistic properties of the space-time asymmetry}}

In this section we focus on specific properties of the
space-time asymmetries such as their group structure and the composition
rules. The properties are explicitly seen in terms of the
scale velocity $\bbox{\nu}$ defined in the way

\begin{equation}
\bbox{\nu} =\frac{\bbox{a}}{\sqrt{1+a^2}}.
\label{eq:u120}
\end{equation}
The scale velocity does not represent state of real motion but
characterizes the state of scale.
In terms of the scale velocity $\bbox{\nu}$, the total energy
(\ref{eq:u97}) of a free particle can be expressed as follows

\begin{equation}
E =\frac{1}{\sqrt{1\!-\!\nu^2}}
\left({\cal E}-\bbox{\nu}\cdot\bbox{p}\right).
\label{eq:u121}
\end{equation}
The dispersion relation (\ref{eq:u97}) can be thus considered as
the Lorenz transformation of energy expressed in terms of the scale
velocity $\bbox{\nu}$. Exploiting the expressions (\ref{eq:u106})
and (\ref{eq:u108}), the
free energy ${\cal E}$ can be rewritten in the way

\begin{equation}
{\cal E}^2 \equiv p^2+m_0^2 = p_{\nu}^2+ m^2
\equiv {\cal E}_{\nu}^2
\label{eq:u122}
\end{equation}
where

\begin{equation}
\bbox{p}_{\nu}= \frac{m}{\sqrt{1\!-\!\nu^2}}\bbox{\nu}.
\label{eq:u123}
\end{equation}
If the motion velocity $\bbox{v}$ is null, the momentum
$\bbox{p}_{\nu}$
corresponds to the minimal energy given by Eq. (\ref{eq:u99}).
Inserting Eqs. (\ref{eq:u106}) and (\ref{eq:u107}) into Eq.
(\ref{eq:u121}) and performing some elementary algebra we get

\begin{equation}
m =\frac{1}{\sqrt{1\!-\!\nu^2}}
\left({\cal E}_{\nu}-\bbox{\nu}\!\cdot\!\bbox{p}_{\nu}\right),
\label{eq:u124}
\end{equation}
\begin{equation}
0 = \bbox{p}_{\it 0} = \bbox{p}_{\nu} +
\frac{\bbox{\nu}(\bbox{\nu}\!\cdot\!\bbox{p}_{\nu})}{\nu^2}
\left(\frac{1}{\sqrt{1\!-\!\nu^2}}-1\right) -
\frac{\bbox{\nu}{\cal E}_{\nu}}{\sqrt{1\!-\!\nu^2}} .
\label{eq:u125}
\end{equation}
These equations represent the Lorenz transformation
$\Lambda(\bbox{\nu})$ of free energy ${\cal E}_{\nu}$ and the
momentum $\bbox{p}_{\nu}$ in dependence on the scale velocity $\bbox{\nu}$.
Because the same is valid for any other velocity $\bbox{\nu}'$,

\begin{equation}
p_{\it 0} = \Lambda(\bbox{\nu}')p_{\nu'},
\label{eq:u126}
\end{equation}
we can write

\begin{equation}
p_{\nu} = \Lambda(-\bbox{\nu})\Lambda(\bbox{\nu}')p_{\nu'}
= R(\bbox{\phi}) \Lambda(\bbox{\nu}'') p_{\nu'} .
\label{eq:u127}
\end{equation}
The matrix $R$ formally describes the Thomas precession around the
axis $\bbox{\phi}=\bbox{\nu}'\times\bbox{\nu}$.
The relativistic transformation (\ref{eq:u127}) connects
the mechanical variables expressed relative
to space-time with different asymmetries.
We conclude from the character of the transformation
that the scale velocities
comply the standard relativistic composition rule

\begin{equation}
\bbox{\nu}'' = \frac{\bbox{\nu}'\sqrt{1\!-\!\nu^2}
+\bbox{\nu}\left[
\left(1\!-\!\sqrt{1\!-\!\nu^2}\right)
\bbox{\nu}\!\cdot\!\bbox{\nu}'/\nu^2
-1\right]}
{1-\bbox{\nu}\!\cdot\!\bbox{\nu}'}.
\label{eq:u128}
\end{equation}
According to the relation (\ref{eq:u120}), this determines the
composition of the corresponding space-time asymmetries $\bbox{a}$
as well.

Consequently, we consider two types of the velocities. First one is
the motion velocity $\bbox{v}$ which determines the state of particle
motion. The velocity represents change of the particle position with
a change
of time in space-time in which the particle is embedded.
The motion velocities are composed according to Eq. (\ref{eq:u50}) and
(\ref{eq:u55}). The second type of velocity is the scale velocity
$\bbox{\nu}$ which characterizes state of scale of the
particle with respect to the state of scale of a reference system.
Structures of both the particle and the reference system are considered
to be fractals of various fractal dimensions.
The reference system can be another particle being a fractal object
or the fractal structure of the (QCD) vacuum itself.
The intimate structure of the later consists of the infinite number of
gluons and quark-antiquark pairs at small scales.

The scale velocities $\bbox{\nu}$ depend on fractal characteristics of
both particle and the reference system  (see the next
section). They
are composed according to Eq. (\ref{eq:u128}) which represents
the special scale relativity law. In this sense, the scale relativity
concerns the transformations of particle characteristics
expressed with respect to self-similar structures which
are fractals of various fractal dimensions. Single fractal
structures have different anomalous fractal dimensions and play
analogous role as the inertial systems in the motion relativity.
The above scale relativistic transformations reflect
the fact that there does not exist an absolute scale reference
system connected either with a fractal object or with any
particular vacuum fractal structure.

The invariants of the scale relativistic transformations
are the scalar products

\begin{equation}
(p_1p_2)_a = \frac{1}{1\!+\!a^2}\left[
E_1E_2-\bbox{p}_1\!\cdot\!\bbox{p}_2
+E_1\bbox{a}\!\cdot\!\bbox{p}_2
+E_2\bbox{a}\!\cdot\!\bbox{p}_1
-(\bbox{a}\!\times\!\bbox{p}_1)(\bbox{a}\!\times\!\bbox{p}_2)
\right].
\label{eq:u129}
\end{equation}
It follows from the previous sections that the scalar products
do not change under the relativistic transformations
(\ref{eq:u84}) concerning the relativity of motion.
Moreover, the scalar products are invariant under the scale
transformations (\ref{eq:u127}) as well.
Really, if we insert Eq. (\ref{eq:u97}) for the energies $E_1$ and $E_2$
into the relation (\ref{eq:u129}), we get

\begin{equation}
(p_1p_2)_a =
{\cal E}_1{\cal E}_2 - \bbox{p}_1\!\cdot\!\bbox{p}_2.
\label{eq:u130}
\end{equation}
The scalar product is expressed in the standard way in terms of free energies
and momenta and is manifestly invariant under the scale
transformations (\ref{eq:u127}). In other words we can write

\begin{equation}
(p_1p_2)_a = (p_1'p_2')_b .
\label{eq:u131}
\end{equation}

We have to answer one more question which is how does the
motion velocity $\bbox{v}$ change with a change of the scale
velocity $\bbox{\nu}$. The answer follows from the Lorenz transformation

\begin{equation}
t =\frac{1}{\sqrt{1\!-\!\nu^2}}
\left({\cal T}+\bbox{\nu}\!\cdot\!\bbox{x}\right),
\label{eq:u132}
\end{equation}

\begin{equation}
\bbox{x}_{\it\!o} = \bbox{x} +
\frac{\bbox{\nu}(\bbox{\nu}\!\cdot\!\bbox{x})}{\nu^2}
\left(\frac{1}{\sqrt{1\!-\!\nu^2}}-1\right) +
\frac{\bbox{\nu}{\cal T}}{\sqrt{1\!-\!\nu^2}}
\label{eq:u133}
\end{equation}
of the coordinates and time with respect to the scale velocity
$-\bbox{\nu}$. The transformation is identical with the scale
transformation of the energy (\ref{eq:u124}) and momentum
(\ref{eq:u125}) except the opposite sign of the scale velocity
$\bbox{\nu}$. According to Eq. (\ref{eq:u120}), this
corresponds to the opposite signs of the asymmetry $\bbox{a}$
in the invariants (\ref{eq:u28}) and (\ref{eq:u96}),
respectively.

The treatment of the symbols here should be taken carefully.
Canonical variables associated with the particle momentum
$\bbox{p}$ and the energy $E$ are the particle position $\bbox{x}$
and the time $t$.
The symbol ${\cal T}$ stands for the time parameter associated
with the particle free energy ${\cal E}$.
We have to realize that rate of time which determines purely
motion is given by the variable $t$.
Exactly this time determinates rate of motion contrary to the
time ${\cal T}$ which is influenced by the rate of scale change.
Therefore, any  motion velocity is given as a derivative with
respect to $t$.
In this sense, the variables

\begin{equation}
\bbox{v} \equiv \frac{d\bbox{x}}{dt}, \ \ \ \ and \ \ \ \
\bbox{v}_{\it\!o} \equiv \frac{d\bbox{x}_{\it\!o}}{dt}
\label{eq:u134}
\end{equation}
are the motion velocities in space-time with and without the
asymmetry $\bbox{a}$, respectively.
Differentiating Eq. (\ref{eq:u133}) with respect to $t$ and using
Eq. (\ref{eq:u132}) one gets

\begin{equation}
\bbox{v}_{\it\!o} = \bbox{v} + \bbox{\nu} +
\frac{\bbox{\nu}(\bbox{\nu}\!\cdot\!\bbox{v})}{\nu^2}
\left(\sqrt{1\!-\!\nu^2}-1\right) .
\label{eq:u135}
\end{equation}
As the relation holds for any asymmetry, the change
of the asymmetry $\bbox{\nu}$ results in the change of the velocity
$\bbox{v}$ in the way which corresponds to the relation

\begin{equation}
x_{\nu} = \Lambda(\bbox{\nu})\Lambda(-\bbox{\nu}')x_{\nu'}
= R(\bbox{\phi}) \Lambda(-\bbox{\nu}'') x_{\nu'} .
\label{eq:u136}
\end{equation}
Exploiting Eq. (\ref{eq:u135}), one can show that

\begin{equation}
m = \frac{\widetilde{m}_0}{\sqrt{1-v_0^2}},
\label{eq:u137}
\end{equation}
where

\begin{equation}
m_0 = \frac{\widetilde{m}_0}{\sqrt{1-\nu^2}} .
\label{eq:u138}
\end{equation}
From the relations we see that the total mass of a particle $m$
is, beside the state of motion, determined also by a state of
scale. If the motion diminishes,
$\bbox{v}_0\rightarrow \bbox{\nu}$, and the mass of the particle $m$
becomes its rest mass $m_0$.
The rest mass is given by a mass $\widetilde{m}_0$ in terms of
the scale velocity $\bbox{\nu}$. If one admits the asymmetry of
space-time fluctuates in dependence on scale,
the rest mass $m_0$ becomes function of the scale
velocity fluctuations. If the fluctuations of $\bbox{\nu}$ have
fractal character, the particle may be identified with
fluctuations of a point-like object with the mass $\widetilde{m}_0$.
The energy $\widetilde{E}$ and the momentum
$\widetilde{\bbox{p}}$ of the point-like object are related to
the energy $E$ and the momentum $\bbox{p}$ of the particle as
follows

\begin{equation}
E = \frac{\widetilde{E}}{\sqrt{1-\nu^2}}, \ \ \ \ \ \ \ \ \
\bbox{p} = \frac{\widetilde{\bbox{p}}}{\sqrt{1-\nu^2}}.
\label{eq:u139}
\end{equation}
Inserting this into equation (\ref{eq:u96}), we get the invariant
in the compact metric form

\begin{equation}
p^2 =
\eta_{\mu\nu}(-\bbox{a})\widetilde{p}^{\mu}\widetilde{p}^{\nu} =
\widetilde{E}^2-\widetilde{\bbox{p}}^{2}+
2\widetilde{E}\bbox{a}\!\cdot\!\widetilde{\bbox{p}}-
(\bbox{a}\!\times\widetilde{\bbox{p}})^2 = m_0^2 .
\label{eq:u140}
\end{equation}

\vskip 0.5cm
{\section {Interactions of fractal objects}}

The ability of fractals to structure space-time was discussed in
Ref. \cite{Nottale}. Such approach gives us possibility to
attribute geometrical notions to the structural parameters
characterizing fractal properties of free particles.
The need to satisfy the principles of the scale-motion relativity
implies replacement of the scale
independent physical laws by the scale dependent equations. This
concerns the energy and momentum which in the presence of the
space-time anisotropy $\bbox{a}$ are converted to the variables satisfying
the dispersion relation (\ref{eq:u97}). Intuitively, the anisotropy of
space-time could be induced in the interaction region at small scales
as a result of the fractality of the interacting constituents.

Consider an ultra-relativistic collision of two fractal objects
(hadrons or nuclei) in their total center-of-energy system $E_1=E_2$.
Suppose the fractals have different anomalous fractal dimensions
$\delta_1<\delta_2$.
According to our working hypothesis,
the interaction of the fractals induces a space-time
asymmetry $\bbox{a}=(0,0,a)$ at small scales.
Due to the asymmetry, momenta of the produced
particles are shifted  against motion of the fractal with
larger anomalous dimension (\ref{eq:u19}).
Here all the momenta are considered to be shifted accordingly.
As the fractal objects collide at  ultra-relativistic
energies, one can neglect their masses and write

\begin{equation}
\sqrt{1\!+\!a^2}p_1-\bbox{a}\!\cdot\!\bbox{p}_1 =
\sqrt{1\!+\!a^2}p_2-\bbox{a}\!\cdot\!\bbox{p}_2 =\sqrt{s}/2.
\label{eq:u141}
\end{equation}
Because the incoming momenta $\bbox{p}_1$ and $\bbox{p}_2$
are alined with the asymmetry
$\bbox{a}$, the relation implies the following formulae

\begin{equation}
\bbox{p}_1+\bbox{p}_2 = \bbox{a}\sqrt{s}, \ \ \ \ \
p_1+p_2 = \sqrt{s}\sqrt{1\!+\!a^2}, \ \ \ \ \ 4p_1p_2 = s.
\label{eq:u142}
\end{equation}
Let us denote the momentum of the recoil particle produced in the
constituent interaction as $\bbox{q}'$.
The corresponding momentum fractions are given in the way

\begin{equation}
\chi_1= \frac{(\bbox{p}_2\!\cdot\!\bbox{q}')_a}
{(\bbox{p}_1\!\cdot\!\bbox{p}_2)_a}, \ \ \ \ \ \
\chi_2= \frac{(\bbox{p}_1\!\cdot\!\bbox{q}')_a}
{(\bbox{p}_1\!\cdot\!\bbox{p}_2)_a},
\label{eq:u143}
\end{equation}
where, according to Eq. (\ref{eq:u130}), the scalar products have
the form

\begin{equation}
(\bbox{p}_i\!\cdot\!\bbox{q}')_a =
p_{i}\sqrt{q^{'2}+m_2^2} -
\bbox{p}_{i}\!\cdot\!\bbox{q}'.
\label{eq:u144}
\end{equation}
When using Eq. (\ref{eq:u142}), (\ref{eq:u144}), and (\ref{eq:u97}),
one can express the sum of the fractions as follows

\begin{equation}
\chi_1 + \chi_2 = \frac{2\sqrt{1\!+\!a^2}
\sqrt{q^{'2}+m_2^2} - 2\bbox{a}\!\cdot\!\bbox{q}'}
{\sqrt{s}} =
\frac{2E'}{\sqrt{s}} .
\label{eq:u145}
\end{equation}
Now we substitute for the fractions $\chi_i$ their expressions
from Eq. (\ref{eq:u13}) and get

\begin{equation}
\sqrt{(1+a^{2})(\mu^{2}_z + \mu^{2}_{\bot})}
-a\mu_z =
\sqrt{\omega^{2}_1+\mu^{2}_1} +
\sqrt{\omega^{2}_2+\mu^{2}_2} -\left(\omega_1-\omega_2\right) .
\label{eq:u146}
\end{equation}
Here we have used the notations

\begin{equation}
\mu_z = \frac{2q_z'}{\sqrt{s}} , \ \ \ \ \
\mu_{\bot} =
\frac{2m_{\bot}'}{\sqrt{s}} , \ \ \ \ \ \
m_{\bot}' = \sqrt{q_{\bot}^{'2}+m_2^2}.
\label{eq:u147}
\end{equation}
The longitudinal  and transversal components of the momentum
$\bbox{q}'$ with respect to the
collision axis are denoted by $q_{z}'$ and $q_{\bot}'$, respectively.
As follows from the conservation of the free energy
(\ref{eq:u101}), Eq. (\ref{eq:u146}) splits into two parts

\begin{equation}
\sqrt{(1+a^{2})(\mu^{2}_z + \mu^{2}_{\bot})} =
\sqrt{\omega^{2}_1+\mu^{2}_1} +
\sqrt{\omega^{2}_2+\mu^{2}_2} ,
\label{eq:u148}
\end{equation}

\begin{equation}
a\mu_z = \omega_1-\omega_2 .
\label{eq:u149}
\end{equation}
The obtained system for the unknown variables $\mu_{z}$ and
$\mu_{\bot}$ depends on the parameter $a$.
The variation range of the variables is given by the
condition $\chi_1+\chi_2\le 1$. According to Eqs. (\ref{eq:u145})
and (\ref{eq:u146}), it can be rewritten as follows

\begin{equation}
(\mu_z-a)^2 + (1\!+\!a^2)\mu^2_{\bot} \le
1\!+\!a^2 .
\label{eq:u150}
\end{equation}
The $\mu_z$ and $\mu_{\bot}$ are bounded inside the
ellipsoid given by the asymmetry $a$.
If we approach the phase-space limit of the reaction
(\ref{eq:u3}), the variables tend to their boundary values
and satisfy the equation of the ellipsoid.
Similar applies for any other particle produced in the elementary
interaction.
The particle's momentum $\bbox{q}'$ and energy $E'$ are
connected by the dispersion relation (\ref{eq:u97})
which can be expressed in the way

\begin{equation}
\left(\frac{q_z'}{E'}-a\right)^2
+ (1\!+\!a^2)\left(\frac{m_{\bot}'}{E'}\right)^2
= 1+a^2 .
\label{eq:u151}
\end{equation}
Exploiting the relations

\begin{equation}
\frac{q_z'}{E'} =
\frac{\sqrt{s}}{2E'}\mu_z =
\frac{\mu_z}{\chi_1+\chi_2} , \ \ \ \ \ \
\frac{m_{\bot}'}{E'} =
\frac{\sqrt{s}}{2E'}\mu_{\bot} =
\frac{\mu_{\bot}}{\chi_1+\chi_2} ,
\label{eq:u152}
\end{equation}
it can be rewritten into the form

\begin{equation}
\left(\frac{\mu_z}{\chi_1+\chi_2}-a\right)^2
+ (1\!+\!a^2)\left(\frac{\mu_{\bot}}{\chi_1+\chi_2}\right)^2
= 1+a^2 .
\label{eq:u153}
\end{equation}
The values of $\mu_z/(\chi_1\!+\!\chi_2)$ are limited
within the interval

\begin{equation}
-a_{-}\le\frac{\mu_z}{\chi_1+\chi_2}\le a_{+}
\label{eq:u154}
\end{equation}
where

\begin{equation}
a_{\pm} = \sqrt{1+a^{2}}\pm a.
\label{eq:u155}
\end{equation}
According to the kinematics of the process, the maximal value
of $\mu^{max}_z/(\chi_1\!+\!\chi_2)=a_{+}$ should correspond
to $\chi_2=0$. The minimal value of
$\mu^{min}_z/(\chi_1\!+\!\chi_2)=-a_{-}$ is given
by $\chi_1=0$.
The maximum of $\mu_{\bot}/(\chi_1\!+\!\chi_2)=1$
should be achieved for
$\chi_1=\chi_2$ and thus for $\mu_z/(\chi_1+\chi_2)=a$.
The conditions are satisfied by the linear combination

\begin{equation}
\mu_z =
(\chi_1+\chi_2)a+(\chi_1-\chi_2)\sqrt{1+a^{2}}.
\label{eq:u156}
\end{equation}
Substituting the expression (\ref{eq:u156}) into the left
side of the relation (\ref{eq:u149}) and using the formulae
(\ref{eq:u13})-(\ref{eq:u15}),
one arrives at the quadratic
equation for the unknown parameter $a$.
Its solution, which complies the physical requirements on the
kinematics of the process, is $a=\bar{a}$. Reminding Eq.
(\ref{eq:u16}),  this explicitly reads

\begin{equation}
a = \frac{\alpha-1}{2\sqrt{\alpha}}\xi
\label{eq:u157}
\end{equation}
where the scale factor  $\xi$ is given by Eq.(\ref{eq:u17}).
Using Eqs. (\ref{eq:u15}), (\ref{eq:u148}), and (\ref{eq:u149}),
one can express the variables $\mu_z$ and $\mu_{\bot}$
in a simple form

\begin{equation}
\mu_z = \mu_1-\mu_2 , \ \ \ \ \ \ \ \
\mu_{\bot} = 2\sqrt{\mu_1\mu_2} .
\label{eq:u158}
\end{equation}
In this view, Eq. (\ref{eq:u156}) is the first of the following
two equations

\begin{equation}
\mu_1\!-\!\mu_2 = \frac{1}{\sqrt{1\!-\!\nu^2}}
\left[(\chi_1\!-\!\chi_2) + \nu (\chi_1\!+\!\chi_2)\right],
\label{eq:u159}
\end{equation}

\begin{equation}
\mu_1\!+\!\mu_2 = \frac{1}{\sqrt{1\!-\!\nu^2}}
\left[(\chi_1\!+\!\chi_2) + \nu (\chi_1\!-\!\chi_2)\right],
\label{eq:u160}
\end{equation}
representing the scale transformation of energy and momentum
along the $z$-axis.
This is seen if we realize that the above combinations of
fractions are expressed in terms of the
corresponding quantities in the way

\begin{equation}
\mu_z = \mu_1-\mu_2 = \frac{2}{\sqrt{s}}q_z' , \ \ \ \ \
\mu_1 + \mu_2= \frac{2}{\sqrt{s}}{\cal E}_{q'} ,
\label{eq:u161}
\end{equation}

\begin{equation}
\chi_z = \chi_1-\chi_2 = \frac{2}{\sqrt{s}}Q_z' , \ \ \ \ \
\chi_1 + \chi_2= \frac{2}{\sqrt{s}}{\cal E}_{Q'}.
\label{eq:u162}
\end{equation}
The scale transformations connect the momentum and the free energy of the
recoil expressed relative to space-time with and
without the asymmetry $\bbox{a}=(0,0,a)$, respectively.
The conservation of the transverse mass

\begin{equation}
\chi_{\bot} = 2\sqrt{\chi_1\chi_2} = 2\sqrt{\mu_1\mu_2} = \mu_{\bot}
\label{eq:u163}
\end{equation}
preserves the invariant forms

\begin{equation}
\mu_z^2+\mu_{\bot}^2 = (\mu_1+\mu_2)^2, \ \ \ \ \ \
\chi_z^2+\chi_{\bot}^2 = (\chi_1+\chi_2)^2
\label{eq:u164}
\end{equation}
which are equivalent to the invariant relations

\begin{equation}
{q'_z}^2+({q'_{\bot}}^{\!2}+m_2^2) = {\cal E}_{q'}^2, \ \ \ \ \ \
{Q'_z}^2+({Q'_{\bot}}^{\!2}+m_2^2) = {\cal E}_{Q'}^2.
\label{eq:u165}
\end{equation}

Similar should apply to the inclusive particle $m_1$. The
only difference is that the inclusive particle is fixed
by its momentum $\bbox{Q}$ while the momentum of the
recoil is  determined from the requirement on minimal
resolution concerning the fractal measure $z$.
The scale transformations of the energy and momentum of the inclusive
particle with the scale velocity $\nu$ oriented along the $z$-axis

\begin{equation}
\tilde{\mu}_1\!-\!\tilde{\mu}_2 = \frac{1}{\sqrt{1\!-\!\nu^2}}
\left[(\lambda_1\!-\!\lambda_2) + \nu (\lambda_1\!+\!\lambda_2)\right]
\label{eq:u166}
\end{equation}

\begin{equation}
\tilde{\mu}_1\!+\!\tilde{\mu}_2 = \frac{1}{\sqrt{1\!-\!\nu^2}}
\left[(\lambda_1\!+\!\lambda_2) + \nu (\lambda_1\!-\!\lambda_2)\right]
\label{eq:u167}
\end{equation}
have the same form as the transformations
(\ref{eq:u159})  and (\ref{eq:u160}).
The momentum fractions

\begin{equation}
\lambda_1 = \sqrt{\tilde{\mu}_1^{2}+\tilde{\omega}_1^{2}}
-\tilde{\omega}_1 , \ \ \ \ \ \
\lambda_2 = \sqrt{\tilde{\mu}_2^{2}+\tilde{\omega}_2^{2}}
+\tilde{\omega}_2 ,
\label{eq:u168}
\end{equation}
are given in terms of

\begin{equation}
\tilde{\omega}_1 = \tilde{\mu}_1a , \ \ \ \ \ \ \ \ \ \
\tilde{\omega}_2 = \tilde{\mu}_2a
\label{eq:u169}
\end{equation}
similarly as their counterparts $\chi_i$.
The value of $a$ is the same as in Eq. (\ref{eq:u157}).
The combinations of fractions

\begin{equation}
\tilde{\mu}_z = \tilde{\mu}_1-\tilde{\mu}_2
= \frac{2}{\sqrt{s}}q_z , \ \ \ \ \
\tilde{\mu}_1 + \tilde{\mu}_2= \frac{2}{\sqrt{s}}{\cal E}_{q} ,
\label{eq:u170}
\end{equation}
and

\begin{equation}
\lambda_z = \lambda_1-\lambda_2 = \frac{2}{\sqrt{s}}Q_z , \ \ \ \ \
\lambda_1 + \lambda_2= \frac{2}{\sqrt{s}}{\cal E}_{Q}
\label{eq:u171}
\end{equation}
are given in terms of the momentum and the free energy of the inclusive
particle expressed relative to space-time with and without the asymmetry
$a$, respectively.
Conservation of the transverse mass

\begin{equation}
\lambda_{\bot} = 2\sqrt{\lambda_1\lambda_2}
= 2\sqrt{\tilde{\mu}_1\tilde{\mu}_2} = \tilde{\mu}_{\bot}
\label{eq:u172}
\end{equation}
and the relations

\begin{equation}
\tilde{\mu}_z^2+\tilde{\mu}_{\bot}^2
= (\tilde{\mu}_1+\tilde{\mu}_2)^2, \ \ \ \ \ \
\lambda_z^2+\lambda_{\bot}^2 = (\lambda_1+\lambda_2)^2
\label{eq:u173}
\end{equation}
have the same form as  Eqs. (\ref{eq:u163}) and (\ref{eq:u164}).
The longitudinal momentum balance reads
\begin{equation}
x_1a_{+}-x_2a_{-} = \tilde{\mu}_z + \mu_z
\label{eq:u174}
\end{equation}
or
\begin{equation}
x_1-x_2 = \lambda_z + \chi_z
\label{eq:u175}
\end{equation}
relative to the scale reference systems
with or without the asymmetry $\bbox{a}=(0,0,a)$, respectively.

All the expressions are given in terms of the asymmetry $a$
(\ref{eq:u157})
which depends on the ratio $\alpha=\delta_2/\delta_1$
of the anomalous fractal dimensions of the colliding objects.
The collisions of the asymmetric fractal objects  are
characterized by the different fractal dimensions and thus with
$a\ne 0$ ($\alpha\ne 1$).
In the considered scenario, it results in creation of a
domain in which the isotropy of space-time is violated.
If $\delta_2=\delta_1$, there is no polarization of space-time
induced by the interaction ($a=0$).
This corresponds to the collisions of the fractals
possessing equal fractal dimensions. Similar applies to the
situation at lower energies where the interacting objects reveal
no fractal-like constituent substructure.
The asymmetry $a$ changes its sign if $\lambda_1\leftrightarrow\lambda_2$ and
$\alpha\leftrightarrow\alpha^{-1}$, i.g. if the interacting fractals
are mutually interchanged.

The fractal limit is achieved in the phase-space limit at any
energy. In this case, the fractions $\lambda_i$ approach
the limiting values which depend on the detection angle $\theta$
of the inclusive particle in the simple way
\begin{equation}
\lambda_1\rightarrow \lambda_1^{L}=\cos^2(\theta/2) ,
\ \ \ \ \ \ \ \ \ \
\lambda_2\rightarrow \lambda_2^{L}=\sin^2(\theta/2) .
\label{eq:u176}
\end{equation}
The limiting values of the longitudinal and the transversal
components of the momentum fractions $\tilde{\mu}$ and $\mu$
read

\begin{equation}
\tilde{\mu}_z^{L} =
\sqrt{\alpha}\cos^2(\theta/2)-
\frac{1}{\sqrt{\alpha}}\sin^2(\theta/2),
\ \ \ \ \ \ \
\tilde{\mu}_{\bot}^{L} = \sin(\theta) ,
\label{eq:u177}
\end{equation}

\begin{equation}
\mu_z^{L} =
\sqrt{\alpha}\sin^2(\theta/2)-
\frac{1}{\sqrt{\alpha}}\cos^2(\theta/2),
\ \ \ \ \ \ \
\mu_{\bot}^{L} = \sin(\theta) .
\label{eq:u178}
\end{equation}
Last equations correspond to angular parameterization of the
ellipsoid of momentum fractions
concerning the inclusive particle and its recoil, respectively.
The momenta of both particles are shifted against the motion of
the fractal object with larger anomalous fractal dimension
$\delta$. The prolonged
form of the ellipsoid means that the momenta of the particles
increase when directed against
relative more resistent fractal structure.

The scale factor $\xi$ (\ref{eq:u17})
is unity in the fractal limit and
the space-time asymmetry  acquires its maximal value

\begin{equation}
a = \frac{\alpha-1}{2\sqrt{\alpha}} .
\label{eq:u179}
\end{equation}
In this case, the expression for the  scale velocity
(\ref{eq:u120}) takes the simple form

\begin{equation}
\nu = \frac{\alpha-1}{\alpha+1}
\label{eq:u180}
\end{equation}
which depends only on the ratio $\alpha=\delta_2/\delta_1$ of the
anomalous fractal dimensions of the colliding objects.
The velocity has its origin in the asymmetry of the
interaction and vanishes in the collisions of objects which
possess equal fractal dimensions.
The scale velocity represents a space-time "drift"
induced by the interaction of the parton fractals.
The quantity represents no real motion but characterizes
local polarization of the (QCD) vacuum.
The velocity depends on the relative state of scale of the reference
systems and satisfies the scale-relativity composition rule

\begin{equation}
\nu' =
\frac{\nu+\nu''}{1+\nu\nu''} ,
\label{eq:u181}
\end{equation}
which results from one dimensional reduction of Eq. (\ref{eq:u128}).
If we insert the expression (\ref{eq:u180}) into this equation, we get

\begin{equation}
\alpha' = \alpha\alpha''.
\label{eq:u182}
\end{equation}
As concerns the fractal limit, the correspondence leads us to make the
following conclusion.
While the composition of scale
velocities follows Einstein-Lorenz law, the composition of the
corresponding ratio of the anomalous fractal dimensions follows
the multiplicative group law.
Last equation takes even more pronounced
form if reduced to a same anomalous fractal dimension, say
$\delta_2$. In that case, Eq. (\ref{eq:u182})  gives

\begin{equation}
\frac{\delta_3}{\delta_1} =
\frac{\delta_2}{\delta_1}
\frac{\delta_3}{\delta_2}.
\label{eq:u183}
\end{equation}

The reduction means that the state of scale of the reference system
possesses natural scaling property consisting in the following.
If studying constituent interactions in the collisions of an
fractal object 1 with a fractal probe 2,
and then in the interactions of the fractal probe 2 with another
fractal object 3,
one arrives at the same properties as if examining the
fractal 1 by mens of the fractal 3. This is again an explicit expression of
the scale relativity in which single fractal structures play
analogous roles as the inertial systems in the motion relativity.

\vskip 0.5cm
{\section{Does asymmetry of space-time contradict with
Michelson's experiment?}}

The concepts considered in the previous section seem to have general
validity at least from the mathematical point of view. However, in
consistence with our physical intuition, we expect
the space-time asymmetry can be induced mainly at small scales
accessible in
ultra-relativistic hadron and nuclear collisions. The question of
its amount is tightly connected with scales we are dealing with.
On the other side, there is no apparent and explicit
reason why the asymmetry should be exactly zero at any scales,
even in large universe.
In general thinking, we show that existence of
any tiny portion of such asymmetries is in principle not ruled out
neither in the famous Michelson's experiment \cite{Michelson} concerning the
interference of light.

The experiment was accomplished by an interferometer having two arms.
Light beam from a light source was divided into two rays, I and II,
traveling perpendicular to each other along the arms. The mirrors
placed on the ends of the spectrometer arms reflected the light
back to the telescope where the rays interfered with each other.
Assuming the apparatus is placed in a region where the
the propagation of light is not isotropic one could expect
existence of a
phase difference $\Delta t$ between the rays I and II which is due
to the anisotropy. When the apparatus is rotated through an angle
of $90^0$, the orientation of the spectrometer arms is
interchanged and the phase difference becomes $-\Delta t$. Such a
rotation of the apparatus should therefore cause a shift of the
interference fringes between the two rays. The experiment,
however, could not find any such effect at all.

We argue below that this experimental fact alone does not imply
absolute absence of any anisotropy in light propagation.
Let us assume there exists a space-time asymmetry $\bbox{a}$
induced by some reasons. One of the reasons we have suggested in
previous section can be connected with fractality at whatever scale.
The asymmetry results in metric changes $\eta_{\mu\nu}(\bbox{a})$
associated with deformation of the circular form of the light
front. The light front becomes an ellipsoid (\ref{eq:u61}) with
the focus in the point the light was emitted (Fig.1). Consider the
Michelson's spectrometer is placed into such anomaly.
The time $t_I$ and $t_{II}$ which the rays take to travel
in spectrometer arms I and II can be expressed as follows

\begin{equation}
t_I   = x_I
\left(\frac{1}{v_1(\phi)} + \frac{1}{v_2(\phi)}\right), \ \ \ \ \
t_{II}= x_{II}
\left(\frac{1}{v_3(\phi)} + \frac{1}{v_4(\phi)}\right).
\label{eq:u184}
\end{equation}
The angle $\phi$ describes orientation of the spectrometer with respect
to the asymmetry $\bbox{a}$. Because of the asymmetry, the
velocities of light propagation $v_i(\phi)$ in different
directions are not equal and depend on the orientation of the
spectrometer. On the other hand spatial distances
(lengths of spectrometer arms) do not depend on the orientation
of the spectrometer. This follows from the known fact
\cite{Moller}
that the spatial geometry is not simply given by the spatial part
$\eta_{ij}$ of the four dimensional metric
$\eta_{\mu\nu}(\bbox{a})$.
The metric tensor $\eta^{\star}_{ij}$ which determines the
spatial geometry is given by

\begin{equation}
\eta^{\star}_{ij} = -\eta_{ij} + \eta^{\star}_i\eta^{\star}_j ,
\label{eq:u185}
\end{equation}
where

\begin{equation}
\eta^{\star}_i = \frac{\eta_{i0}}{\sqrt{\eta_{00}}} .
\label{eq:u186}
\end{equation}
In the case of the four-dimensional metric (\ref{eq:u29}),
the spatial metric reads

\begin{equation}
\eta^{\star}_{ij} = (1\!+\!a^2)\delta_{ij}.
\label{eq:u187}
\end{equation}
Therefore, the distance

\begin{equation}
d = \sqrt{\eta^{\star}_{ij}x^ix^j}
= \sqrt{1\!+\!a^2}x
\label{eq:u188}
\end{equation}
is invariant under space rotations.
Let us now exploit the following geometrical property of the
velocity ellipsoid (\ref{eq:u61}).
While the sections $v_i(\phi)$ connecting any
point of the ellipsoid with the focus depend on their
orientation $\phi$, the combination

\begin{equation}
\frac{1}{v_1(\phi)} + \frac{1}{v_2(\phi)}
= \frac{2a_{\parallel}}{a_{\bot}^2} = 2\sqrt{1\!+\!a^2}
\label{eq:u189}
\end{equation}
is rotationally invariant i.e. does not depend on the angle
$\phi$. The symbols $a_{\parallel}$ and $a_{\bot}$ denote the big
and small semi-axis of the ellipsoid, respectively.
After inserting expressions (\ref{eq:u188}) and (\ref{eq:u189}) into
Eq. (\ref{eq:u184}),  we get

\begin{equation}
t_I   = 2d_I, \ \ \ \ \ \
t_{II}= 2d_{II}.
\label{eq:u190}
\end{equation}
The relations connect time the light rays take to travel in the
spectrometer arms with the lengths of the arms. Both physical
quantities are expressed in space-time with the asymmetry $\bbox{a}$
giving the relations which are rotational invariant.
Therefore, rotation of the spectrometer apparatus should not
cause a shift of the interference fringes even for
$\bbox{a}\ne 0$.

We could also perform a "gedanken" experiment with tree mirrors which
reflect the rays of light along a triangle $ABC$. Consider the
triangle depicted by the full lines in Fig.2a. Suppose a light signal
is emitted in the point $A$ and travels along
the path $d_1$, $d_2$, and $d_3$.  The corresponding time
interval

\begin{equation}
t_{ABC}   =
\frac{x_1}{v_1(\phi)} + \frac{x_2}{v_2(\phi)}
+ \frac{x_3}{v_3(\phi)}
\label{eq:u191}
\end{equation}
is function of the velocities $v_1(\phi)$, $v_2(\phi)$, and $v_3(\phi)$.
The velocities are shown on the velocity diagram in Fig.2b.
They depend on the orientation $\phi$ of the three
mirrors setup relative to the asymmetry $\bbox{a}$.
If the experimental setup begins
to rotate, the values of $v_i(\phi)$ will change but the value of
$t_{ABC}$ remains invariant with respect to such rotations.
The invariance follows from specific
geometrical properties of any rotational ellipsoid which we outline below.

Let us denote the internal angles of the triangle $ABC$ as
$\alpha_1$,  $\alpha_2$, and $\alpha_3$. They comply the
elementary geometrical property written in the way

\begin{equation}
\frac{d_1}{\sin\alpha_1}
=\frac{d_2}{\sin\alpha_2}
=\frac{d_3}{\sin\alpha_3} \equiv \sqrt{1\!+\!a^2}x_{ABC},
\label{eq:u192}
\end{equation}
where $d_i$ are the opposite sides of the triangle.
The angles among the corresponding velocities  are
denoted by $\beta_1$, $\beta_2$, and $\beta_3$ and shown in Fig.2b.
In a specific mirror setup, they are fixed by the relation

\begin{equation}
\beta_i = \pi - \alpha_i,\ \ \ \ \ \ \ i = 1,2,3.
\label{eq:u193}
\end{equation}
The angles $\beta_i$ do not depend on the orientation $\phi$
and thus do not depend on the rotation of the apparatus as the
whole.
Using the above formulae, Eq. (\ref{eq:u191}) takes the form

\begin{equation}
t_{ABC}   = x_{ABC}\left(
\frac{\sin\beta_1}{v_1(\phi)} + \frac{\sin\beta_2}{v_2(\phi)}
+ \frac{\sin\beta_3}{v_3(\phi)}\right).
\label{eq:u194}
\end{equation}
One can convince itself that there exists following geometrical
property of the rotational ellipsoids. Consider an ellipse which represents a
section of the ellipsoid with a plane passing though the focus of
the ellipsoid. The  focus is common for this ellipse and the ellipsoid.
Let us denote by $v_1(\phi)$, $v_2(\phi)$, and $v_3(\phi)$
the sections connecting three different
points of the ellipse with the common focus.  While magnitudes of
the sections $v_i(\phi)$ depend on the orientation $\phi$ of the ellipsoid,
the combination

\begin{equation}
\frac{\sin\beta_1}{v_1(\phi)} + \frac{\sin\beta_2}{v_2(\phi)}
+ \frac{\sin\beta_3}{v_3(\phi)}
= \frac{2a_{\parallel}}{a_{\bot}^2}
\left(\sin\beta_1+\sin\beta_2+\sin\beta_3\right).
\label{eq:u195}
\end{equation}
remains invariant under any rotation of the ellipsoid.
The symbols $a_{\parallel}$ and $a_{\bot}$ denote the big
and small semi-axis of the ellipsoid, respectively.
Exploiting Eqs. (\ref{eq:u61}) and
(\ref{eq:u192}) - (\ref{eq:u195}), one arrives at the relation

\begin{equation}
t_{ABC}   = d_1 + d_2 + d_3
\label{eq:u196}
\end{equation}
which does not depend on the orientation $\phi$.
Therefore, arbitrary rotation of a three mirror setup will not
cause a shift of the interference fringes of light for
$\bbox{a}\ne 0$.

Let us now imagine a light signal traveling along the
triangles $ABC$ and $ACD$ depicted in Fig.2a in the following order.
Suppose the signal is emitted in the point $A$, then travels along the lines
$d_1$, $d_2$, and $d_3$. The signal is reflected in the
point $A$ back
and travels the distances $d_4$, $d_5$, and $d_6$. The light takes
to travel the whole path during the time

\begin{equation}
t_{ABC} + t_{ACD}  = d_1 + d_2 + d_3 + d_4 + d_5 + d_6 =
t_{out}+t_{int}.
\label{eq:u197}
\end{equation}
As follows the above considerations, this relation does not depend on
the particular choice of the angle $\phi$.
Here we have denoted by $t_{int}$ the time the light ray travels
along the internal line $CA$ to and fro.
According to Eq. (\ref{eq:u189}), the relation between the internal
time $t_{int}$
and the distance $d_3+d_4$ is invariant under space rotations.
Consequently, we get the formula

\begin{equation}
t_{out} = d_1 + d_2 + d_5 + d_6
\label{eq:u198}
\end{equation}
which does not depend on the space-time
asymmetry $\bbox{a}$.
It is possible to think of various trajectories from the
point $A$ to the point $B$.
One of the trajectories is line connecting the points
$ADCB$. A light signal traveling along this
trajectory takes the time

\begin{equation}
t_{tr} = d_{tr} + \Delta_{AB}
\label{eq:u199}
\end{equation}
where

\begin{equation}
d_{tr} = d_6 + d_5 + d_2, \ \ \ \ \ \
\Delta_{AB} = d_1-\frac{x_1}{v_1}
\label{eq:u200}
\end{equation}
are the length of the trajectory and a specific time difference,
respectively.  The time difference $\Delta_{AB}$ is due to the
space-time asymmetry $\bbox{a}$ but does not depend on the shape
and length of the trajectory $ADCB$. Therefore, the light signal
takes the same time $t_{tr'}=t_{tr}$ when traveling along another
trajectory $AD'C'B$ with the same length $d_{tr'}=d_{tr}$.
If the trajectory becomes more complicated its length increases
so that
$d_{tr}>>\Delta_{AB}$. In the fractal limit one gets the
asymptotic relation

\begin{equation}
t_{tr} = d_{tr}
\label{eq:u201}
\end{equation}
which does not depend on any particular value of the
space-time asymmetry $\bbox{a}$ corresponding to a specific
resolution.

\vskip 0.5cm
{\section {Summary}}

The questions addressed in the paper concern ultra-relativistic
nuclear interactions at constituent level.
They are connected with the notions such as locality,
self-similarity, fractality and the scale relativity.

We have discussed some aspects of the relation between
the fractality of the interacting objects and the properties
of space-time induced by their interactions.
The relation is relevant for small scales where the parton composition
of the hadron objects is supposed to reveal a fractal-like substructure.
The assumption has fundamental consequence which is breaking
of the reflection invariance in dependence on scale.
We have elaborated the formalism concerning the special relativity
in space-time with broken reflection invariance.
Our treatment accounts for change in the dispersion relation
including change of the metrics in space-time.
If we ignore quantum uncertainties in the energies and momenta we
obtain explicit relations between the energy/momentum and
the velocity in space-time characterized by the asymmetry
$\bbox{a}$. The asymmetry was shown to be a relative quantity
governed by scale relativistic principles. This concerns Lorenz invariance
with respect to the scale velocity $\bbox{\nu}$ connected with the
asymmetry $\bbox{a}$. If considering the asymmetry as fluctuating
intrinsic property of space-time itself,
local structure of the quantity
should depend on scale and the relativistic invariants only.
This underlines application of the functional self-similarity
to the expression

\begin{equation}
\bbox{a}[\xi, \tau^2] =
\bbox{a}[\xi,
t^{2}-\bbox{x}^{2}-
2t\bbox{a}\!\cdot\!\bbox{x} - (\bbox{a}\!\times\!\bbox{x})^2] .
\label{eq:u202}
\end{equation}
In this vision, the asymmetry should depend on a scale parameter $\xi$ and
through invariants of motion on the space-time
positions and its own values at another scales.
Adequate description one has to  search for is, in our opinion,  within the
renorm group approach to the self-similar systems.
This should include elementary quantum fields as
source of the space-time fluctuations $\bbox{a}$.
In view of our results, increase of stochasticity
of the space-time asymmetry with decreasing scales would result
in fractal-like motion of "elementary" point-like
objects with respect to their momenta.
This implies change of their rest mass $m_0$ in dependence on the
value of $\bbox{\nu}$.

In the paper we have considered space-time asymmetry $a$
and have supposed it can be induced in the
ultra-relativistic nuclear collisions.
The asymmetry was expressed by the anomalous fractal
dimensions of the colliding objects.
The fractal dimensions characterize hadronic constituent substructure
revealed at high energies.
The relation is illuminated with respect to maximal
resolution in the definition of the scaling variable $z$.
Existence of even tiny portion of space-time asymmetry at any
scales have been discussed within the framework of the
Michelson's experiment.
We have shown that such asymmetry is in principle
not ruled out by similar experiments concerning interference of
light signals.

Presented approach to the $z$ scaling shows that the observed
regularity can have relevance to fundamental principles of
physics at small scales.
The general assumptions and ideas discussed here underline
need of searching for new approaches to physics at
ultra-relativistic energies.
This concerns better understanding of scale dependence of physical
laws in the domain tested by large accelerators of hadrons and nuclei.

\vskip 1cm

{\Large \bf Acknowledgment}

\vskip 0.5cm

This work has been partially supported
by the grant No. 020475 of the Czech Ministry of Education Youth
and Physical Training.

\newpage
\vskip 1cm
{\section{Appendix}}

\vskip 0.5cm

Here we present derivation of the relativistic
transformations in one dimensional case in detail.
Without any loss of generality, the linearity of the
transformations can be expressed as follows

\begin{equation}
x'  = \gamma (u)[x-ut] ,
\label{eq:u203}
\end{equation}

\begin{equation}
t'  = \gamma (u)[A(u)t-B(u)x] ,
\label{eq:u204}
\end{equation}
where $\gamma$, $A$, and $B$ are unknown functions of a velocity
$u$. Let us compose the transformation with the successive one

\begin{equation}
x''  = \gamma (v')[x'-v't'] ,
\label{eq:u205}
\end{equation}

\begin{equation}
t''  = \gamma (v')[A(v')t'-B(v')x'] .
\label{eq:u206}
\end{equation}
The result can be written in the form

\begin{equation}
x''  = \gamma(u)\gamma(v')[1+B(u)v']
\left[x-\frac{u+A(u)v'}{1+B(u)v'}t\right] ,
\label{eq:u207}
\end{equation}

\begin{equation}
t''  = \gamma(u)\gamma(v')[A(u)A(v')+B(v')u]
\left[t-\frac{A(v')B(u)+B(v')}{A(u)A(v')+B(v')u}x\right] .
\label{eq:u208}
\end{equation}
The principle of relativity is expressed by the
group structure of the transformations. The
condition tells us that Eqs. (\ref{eq:u207}) and (\ref{eq:u208}) keep
the same form as the initial ones in terms of the composed
velocity

\begin{equation}
v  = \frac{u+A(u)v'}{1+B(u)v'} .
\label{eq:u209}
\end{equation}
The requirement can be satisfied under the following
conditions

\begin{equation}
\gamma(v)  = \gamma(u)\gamma(v')[1+B(u)v'] ,
\label{eq:u210}
\end{equation}

\begin{equation}
\gamma(v)A(v)  = \gamma(u)\gamma(v')[A(u)A(v')+B(v')u] ,
\label{eq:u211}
\end{equation}

\begin{equation}
\frac{B(v)}{A(v)} = \frac{A(v')B(u)+B(v')}{A(u)A(v')+B(v')u} .
\label{eq:u212}
\end{equation}
The isotropy of space-time results in
the requirement that the change of orientations of the variable axis are
indistinguishable, provided $u'=-u$. This leads to the parity
relations $\gamma(-u)=\gamma(u)$, $A(-u)=A(u)$, and
$B(-u)=-B(u)$. The relations are sufficient for the derivation
of the Lorenz transformation.
The theory of relativity tells us that the velocity of a
physical object can not exceed the value
of $c=1$, the velocity of light in the vacuum.
The expression of this statement is the Lorenz transformation which
yields the limitation of any velocity.

If we leave out the constraint on the
reflection invariance,
the unknown functions $\gamma$, $A$, and $B$
do not obey the parity relations resulting from the
isotropy requirement.
Let us combine Eqs. (\ref{eq:u209}), (\ref{eq:u210}), and
(\ref{eq:u211}) into the expression

\begin{equation}
A\left( \frac{u+A(u)v'}{1+B(u)v'}\right) =
\frac{A(u)A(v')+B(v')u}{1+B(u)v'} .
\label{eq:u213}
\end{equation}
Its solution has the form

\begin{equation}
A(u) = 1 - 2au ,
\label{eq:u214}
\end{equation}
provided $B(u)v' = B(v')u$.
The condition gives the function $B(u) = u$ with the
normalization constant $c$ included already in the
definition of the variable $u$.
The solution satisfies Eq. (\ref{eq:u212}) as well.
The violation of the space-time reflection invariance is
expressed by a non-zero value of the parameter $a$.
In terms of $a$, the composed velocity (\ref{eq:u209})
can be written as follows

\begin{equation}
v  = \frac{v'+u-2auv'}{1+uv'} .
\label{eq:u215}
\end{equation}
Using this relation, Eq. (\ref{eq:u210}) becomes

\begin{equation}
\gamma\left(\frac{v'+u-2auv'}{1+uv'}\right)
= \gamma(u)\gamma(v')(1+uv') .
\label{eq:u216}
\end{equation}
Its solution, which for $a=0$ is given by the standard $\gamma$
factor, has the form

\begin{equation}
\gamma(u) = \frac{1}{\sqrt{1-2au-u^{2}}} .
\label{eq:u217}
\end{equation}
The detailed classification of the linear transformations
(\ref{eq:u203}) and (\ref{eq:u204}) in 1+1 dimensions
was performed in Ref. \cite{Lalan}.

\newpage
\begin{center}

\begin{figure}
\epsfig{file=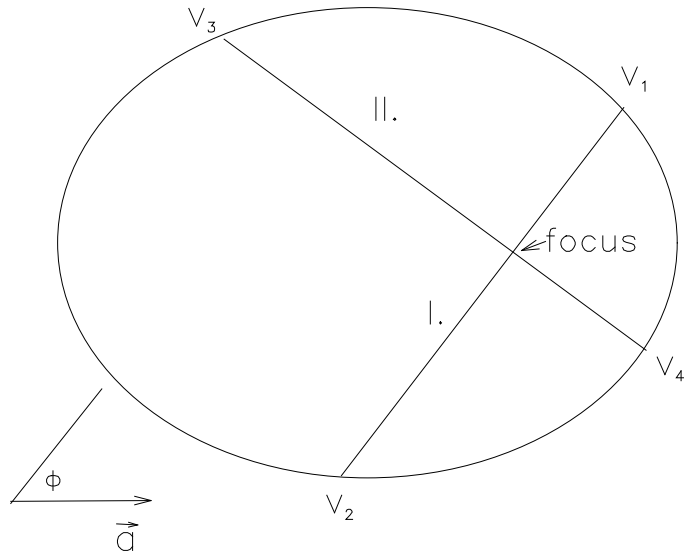, width=90mm}
\vspace*{5mm}
\caption{The velocity diagram in space-time with the asymmetry
$\bbox{a}$. The lines I. and II. correspond to the orientation
of the spectrometer arms in the Michelson's experiment.
}
\end{figure}

\newpage
\begin{figure}
\epsfig{file=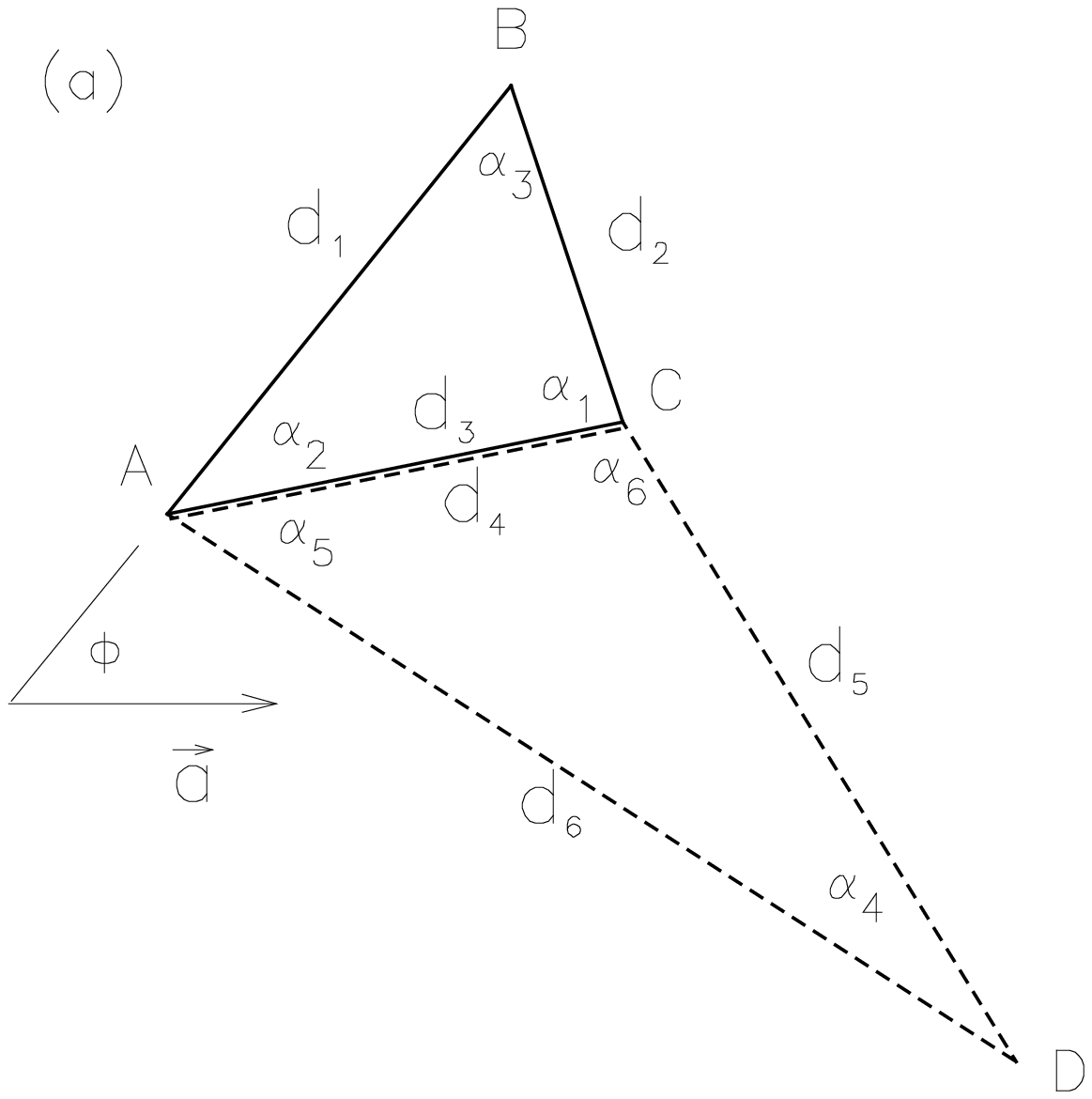, width=90mm}
\end{figure}
\vspace*{10mm}
\begin{figure}
\epsfig{file=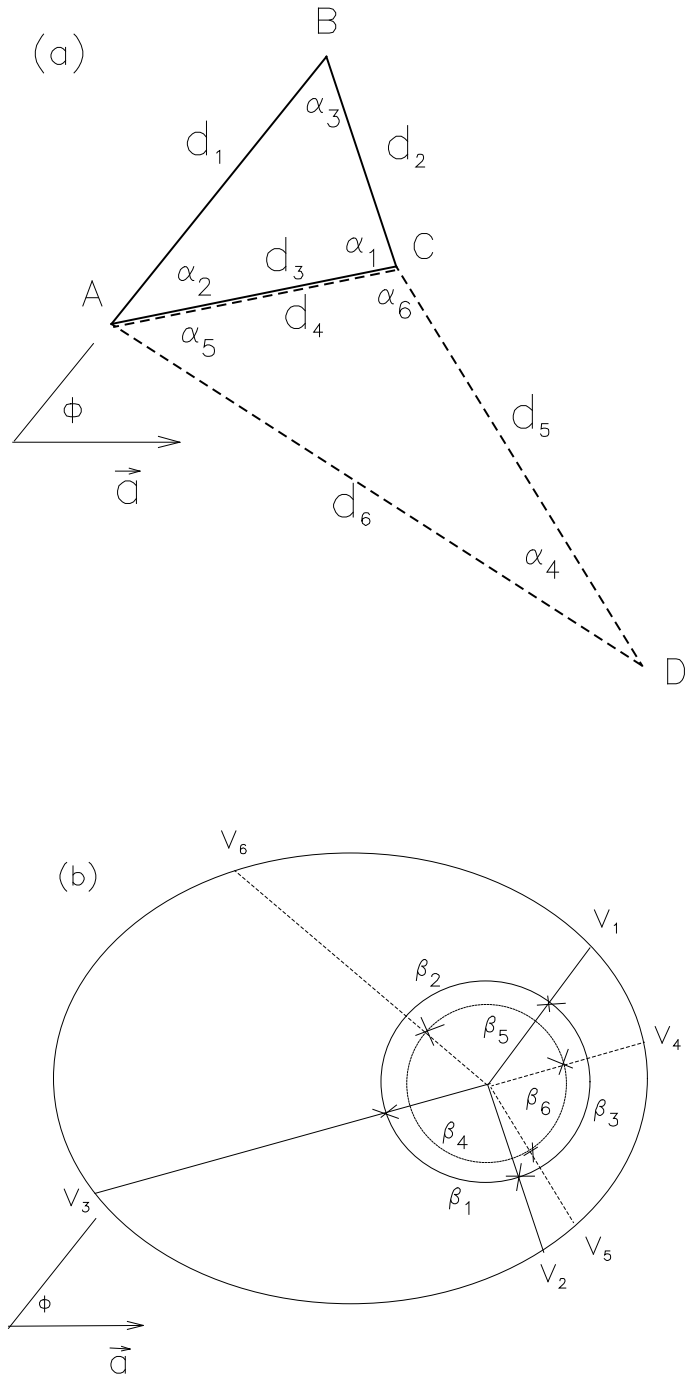, width=90mm}
\vspace*{10mm}
\caption{
(a) The space diagram of a multi-mirror setup. The mirrors
are considered in the points $A$, $B$, $C$, and $D$ reflecting
the light signal along the sketched lines.
(b) The velocity diagram corresponding to the mirrors
    setup shown in Fig.2a.
}
\end{figure}
\end{center}
\end{document}